\documentclass[british,amssymb,aps,prd,twocolumn,amsmath,floats,floatfix,superscriptaddress,nofootinbib,%
               showkeys,showpacs,groupedaddress]{revtex4-1}
\usepackage{amssymb}
\usepackage{amsmath}
\usepackage{verbatim}
\usepackage{mathrsfs}
\usepackage{amsfonts}
\usepackage{latexsym}
\usepackage{epsfig}
\usepackage{color}
\usepackage[usenames,dvipsnames]{xcolor}
\usepackage{graphicx,subfigure}
\usepackage{units}
\usepackage{envmath}
\usepackage{natbib}
\usepackage{ctable}
\usepackage{soul}
\usepackage[unicode=true,pdfusetitle,
 bookmarks=true,bookmarksnumbered=false,bookmarksopen=false,
 breaklinks=false,pdfborder={0 0 1},backref=false,colorlinks=false]{hyperref}
\usepackage{mathtools}
\begin{document}
\definecolor{orange}{rgb}{0.9,0.45,0}
\def\CovDev{D}
\def\Res{{\mathcal R}}
\def\Gammaflat{\hat \Gamma}
\def\metricflat{\hat \gamma}
\def\Dflat{\hat {\mathcal D}}
\def\part_n{\partial_\perp}
\def\Lie{\mathcal{L}}
\def\A{\mathcal{X}}
\def\Aphi{\A_{\phi}}
\def\hAphi{\hat{\A}_{\phi}}
\def\E{\mathcal{E}}
\def\Ham{\mathcal{H}}
\def\M{\mathcal{M}}
\def\R{\mathcal{R}}
\def\p{\partial}
\def\hg{\hat{\gamma}}
\def\hA{\hat{A}}
\def\hD{\hat{D}}
\def\hE{\hat{E}}
\def\hR{\hat{R}}
\def\hcA{\hat{\mathcal{A}}}
\def\hDelt{\hat{\triangle}}
\def\na{\nabla}
\def\dif{{\rm{d}}}
\def\non{\nonumber}
\newcommand{\erf}{\textrm{erf}}
\renewcommand{\t}{\times}
\long\def\symbolfootnote[#1]#2{\begingroup%
\def\thefootnote{\fnsymbol{footnote}}\footnote[#1]{#2}\endgroup}
\title{Perturbation Spectra of Warm Inflation in $f(Q, T)$ Gravity}

 \author{Maryam Shiravand$^{1}$}\email{ma\_shiravand@sbu.ac.ir}
 \author{Mehrdad Farhoudi$^{1}$}\email{m-farhoudi@sbu.ac.ir}
 \author{Parviz Goodarzi$^{2}$}\email{parviz.goodarzi@abru.ac.ir}
 \affiliation{ $^{1}$Department of Physics,
                     Shahid Beheshti University, Evin, Tehran 19839,
                     Iran \\
  $^{2}$Department of Physics, Faculty of Basic Sciences, Ayatollah Boroujerdi University, Boroujerd,
 Iran}

\date{July 10, 2024}
\begin{abstract}
\noindent
 We investigate the warm inflationary scenario within the
context of the linear version of $f(Q,T)$ gravity, coupled with
both the inflaton scalar field and the radiation field, under the
conditions of the strong dissipation regime. First, we calculate
the modified Friedmann equations and the modified slow-roll
parameters. Subsequently, we apply the slow-roll approximations to
derive the scalar power spectrum and the tensor power spectrum.
Also, we develop formulations of the scalar and tensor
perturbations for the $f(Q,T)$ gravity with the warm inflation
scenario. Furthermore, we scrutinize two different forms of the
dissipation coefficient, a constant and a function of the inflaton
field, to determine the scalar spectral index, the
tensor-to-scalar ratio and the temperature for the power-law
potential case. By imposing some constraints on the free
parameters of the model, we attain results in good agreement with
both the Planck $2018$ data and the joint Planck, BK$15$ and BAO
data for the tensor-to-scalar ratio, and consistent results
aligned with the Planck $2018$ data for the scalar spectral index.
In addition, the obtained results are within the range of
observational data for the amplitude of the scalar power spectrum.
Consequently, we are able to revive the power-law potential that
was previously ruled out by observational data. Moreover, for both
dissipation coefficients, the model leads to a scalar spectral
index with the blue and red tilts in agreement with the WMAP three
years data.
\end{abstract}

\maketitle

\section{Introduction}
In recent years, numerous scientific investigations, both
theoretical and observational, have been carried out to comprehend
the dynamics of the Universe. The outcomes of these studies align
closely with the widely accepted cosmological model known as
$\Lambda$CDM, which is based on general relativity (GR).
Concurrently, surveys examining the cosmic microwave background
(CMB) radiation have provided crucial insights into the formation
and development of the Universe~\cite{Hinshaw2013, Aghanim2021}.
Despite the overall agreement, certain perplexing concepts like
the flatness and horizon problems persist within the standard
cosmological model. The concept of inflation has proven to be a
highly successful framework for addressing the issues in the
standard cosmology. It also serves as a foundation for explaining
the generation and evolution of the primordial seeds that give
rise to the large-scale structure of the
Universe~\cite{starobinsky1980, Guth1981, Albrecht1982, linde1982,
linde1990}.

The conventional inflation model consists of two distinct
evolutionary phases. The first phase is characterized by an
accelerated expansion driven by an inflaton field that has
negligible kinetic energy compared to its potential energy, a
condition known as slow-roll. Notably, the quantum fluctuations of
the inflaton can lead to a period of inflation and provide an
explanation for the origin of large-scale
structures~\cite{Barrow1990, Barrow2006}. In simpler terms, cosmic
inflation produces density perturbations, which can be observed
through the measurement of temperature variations in the
CMB~\cite{wmap2003, Tegmark2004}. On the other hand, a broad range
of potential functions capable of describing an inflationary
period have been thoroughly investigated and constrained using
measurements of CMB anisotropies~\cite{martin2014}. In the
subsequent stage, known as reheating, the inflaton undergoes decay
and gives rise to matter and radiation fields similar to a hot big
bang scenario~\cite{Bassett2006}. The major challenge lies in
successfully bridging the evolution of the Universe towards the
end of this era. One intriguing approach to address this issue
involves investigating inflation in the context of the warm
inflation scenario. Warm inflation takes into account thermal
effects and interactions between the inflaton field and other
particles in the Universe~\cite{Bererafang1995, Berera1995,
Berera1997,Berera2023}. These interactions facilitate the
continuous transfer of energy from the inflaton to the surrounding
particles and create a significant thermal environment. The
inclusion of thermal effects introduces dissipation processes and
has the potential to produce noticeable deviations from the
expected outcomes of cold inflation~\cite{Berera1999}. This is
achieved by introducing a dissipation
coefficient~\cite{Berera2023, Oliveira1998, Berera2000, Hall2004}.
The dissipation term can be derived using quantum field theory
methods within a two-stage mechanism for the field
interaction~\cite{Bastero2011, Bastero2013}. Consequently, warm
inflation offers an alternative explanation for the emergence of
cosmic structure. Thus, the advantages of warm inflation encompass
the automatic reheating process at the end of inflation, where the
thermal bath begins to surpass the vacuum energy, as well as the
suppression of contributions to the tensor-to-scalar
ratio~\cite{Bellini1998}. This feature can revive many
inflationary models from the perspective of observational data.

A consequence of the friction term, which characterizes the decay
of the inflaton field into the radiation field during inflation,
is that the density perturbations are now mainly influenced by
larger thermal fluctuations rather than caused by quantum
fluctuations~\cite{Taylor2000}. For scalar perturbations, the
interaction between the inflaton field and the radiation field
leads to entropy perturbations along with adiabatic ones. This
arises from the characterization of warm inflation as an
inflationary model with two fundamental fields. In this framework,
dissipative effects have the capacity to produce a diverse range
of spectra~\cite{Herrera}. In the large-scale limit of slow-roll
conditions, curvature perturbations remain constant and entropy
perturbations become negligible~\cite{Hall2004}. In the
literature, there are many works on cosmological perturbations,
see, e.g.,
Refs.~\cite{Hall2004,Taylor2000,Mukhanov1992,Oliveira2001,
Brandenberger,Bartolo2004,Weinberg, Bari2019, Antonio2022}.

On the other hand, while GR has been successful in predicting
cosmological phenomena (see, e.g.,
Refs.~\cite{Einestein1916,Will:2005va}), it is~not ideal for
explaining the existence of dark sectors (dark matter and dark
energy) and their impacts on the dynamics of the Universe, which
align well with observational data~\cite{Huterer1999, Alam2004,
Alam2007,Ishak2019, Ferreira:2019xrr}. Such shortcomings have
prompted researchers to explore alternative theories of gravity,
see, e.g., Refs.~\cite{Farhoudi2006,farb, Sotiriou2010,
Felice2010, zare1, zare3, Haghani2018,zare5}. Various modified
theories of gravity have also been investigated in the context of
inflationary cosmology, with the aim of providing better alignment
between their predictions and observational
data~\cite{starobinsky1980, Vilenkin1985, Majic1986, Fuji2003,
Carrol2005, Amendola2007, Nojiri2011,
Shabani2014,Saba2018,Rasouli2018,Odintsov2023}.

Alternative approaches in modifying GR involve exploring
geometries beyond the Riemannian framework, such as torsion and
nonmetricity, which provide more geometric descriptions of
gravity. Early explorations in this direction were conducted by
Weyl, whose goal was to establish a geometric interpretation for
electromagnetism and a unified theory encompassing gravitation and
electromagnetism~\cite{Weyl1918, Wheeler2018}. To incorporate this
extension of geometry, Weyl introduced a compensating vector field
and also proposed that during a parallel transformation of a
vector, not~only its direction but also its magnitude undergoes a
change. Within this concept, the connection of the system can be
divided into two parts, such that one part is a connection that
accounts for the variation in the magnitude of the vector, while
the other part is the usual Levi-Civita connection that represents
changes in the direction of the vector under a parallel transport.
An important point is that both connections are torsionless within
the framework of the Weyl geometry. Weyl attempted to interpret
the introduced vector field as the electromagnetic potential, but
his theory ultimately proved unsuccessful and was even rejected by
himself~\cite{Weyl1950}.

Nevertheless, Dirac made an attempt to revive the Weyl geometry.
He demonstrated~\cite{Dirac1973} that the distinction between the
Weyl and Riemannian geometries lies in the expression of the
partial derivative. In fact, through the Weyl gauge
transformation, the Weyl space can be reduced to a Riemannian
space with a metric that is conformally related to the original
metric. However, it has been argued~\cite{Dirac1973} that the Weyl
theory fails to describe the electromagnetic interaction because
the introduced vector field does~not interact with the spinor,
unlike the electromagnetic potential. While the Weyl theory had
limited physical success, it nevertheless presented some
intriguing aspects. Notably, one of the significant features of
the Weyl geometry is the presence of a nonzero covariant
derivative of the metric tensor, which geometrically leads to the
introduction of a quantitative term known as nonmetricity tensor.
To simplify the geometric formalism of gravitation, an alternative
theory called the teleparallel equivalent of GR (TEGR) was
proposed. TEGR utilizes the Weitzenb\"{o}ck connection, which has
zero curvature and nonmetricity tensors but non-zero
torsion~\cite{Maluf2013}.

Another alternative framework called symmetric teleparallel
gravity has also been developed, utilizing a connection with zero
curvature and torsion tensors while incorporating a nonmetricity
tensor to describe gravitational interactions. This symmetric
teleparallel gravity was further modified to coincide with both GR
and $f(Q)$ gravity\rlap,\footnote{The $Q$ term is the nonmetricity
scalar described in Sec.~III.}\
 specifically in a gauge
known as the coincident gauge, where the connection
vanishes~\cite{Naster1999, Jimenez2018}. A significant aspect of
this theory is its ability to separate gravitational and inertial
effects, something not achievable in GR. The symmetric
teleparallel gravity offers an alternative geometrical description
of gravity that is dynamically equivalent to GR. The theory is
governed by the Einstein-Hilbert action in the absence of boundary
terms and naturally gives rise to self-accelerating cosmological
solutions in the early and late stages of the Universe, see, e.g.,
Refs.~\cite{Lu2019, Jimenez2020, Dimakis2022}. For a review of the
teleparallel gravity, see Ref.~\cite{Bahamonde}.

An alternative extension of the symmetric teleparallel gravity
theory involves the coupling between the geometry and the trace of
the energy-momentum tensor in action as an arbitrary function
$f(Q, T)$. This construction enables a straightforward description
of gravitational interactions in the presence of geometry-matter
coupling. The obtained cosmological solutions can account for both
the accelerating and decelerating evolutionary phases of the
Universe. Consequently, $f(Q, T)$ gravity shows promise in
providing valuable insights into describing both the early and
late stages of the Universe~\cite{Xu2019}. To date, numerous works
and several cosmological aspects of $f(Q, T)$ gravity have been
investigated~\cite{Arora:2020tuk,Xu2020,Arora2020,Arora:2021jik,
Gadbail:2021fjf, Gadbail2021,Arora:2021met,Godani2021,Pradhan2021,
Agrawal:2021rur,Pati2021,2022Arora,Pati:2021zew, Shiravand2022,El
Bourakadi2023,Bhagat2023, Tayde2023, Pati2022, Mandal2022,
Narawada2023, Das2023}. In this research, we investigate the warm
inflation in $f(Q, T)$ gravity. Warm inflation has already been
considered in different areas and in alternative gravitational
theories, see, e.g.,
Refs.~\cite{Herrera,Nozari,Setare2012,Jamil2015,Sharif2016,Sharif2017,Amaek2022,Shahid2023,
Yeasmin2023}. Our focus lies in the context of linear version of
$f(Q, T)$ gravity, where in our previous
paper~\cite{Shiravand2022}, we showed that the evolution of a
general linear version of $f(Q,T)$ gravity is~not fixed on a
de~Sitter expansion or on the Minkowski metric. Although, it is
also claimed that the linear version of $f(Q, T)$ gravity is
dynamically equivalent to the linear version of $f(R,T)$ gravity
that is claimed to be GR with a little change in the content of
matter~\cite{Fisher2019,HMoraes,Fisher2020}. However, this
equivalence is a matter of debate, and besides, it provides
another geometrical formalism. In addition, we
justified~\cite{Shiravand2022} the utilization of the linear
version of $f(Q,T)$ gravity by providing a comparison showing
possible differences with what is obtained from the GR case.
Indeed, we indicated that the parameters in the linear version of
$f(Q,T)$ gravity provide better fittings to the observational data
than GR. Furthermore, through this version of gravity and based on
the joint Planck, BK$15$ and BAO data, we
revived~\cite{Shiravand2022} the previously excluded natural
potential.

The work structure is as follows. In Sec.~II, we present some
essential relations on the conventional warm inflation and the
related slow-roll conditions and parameters. Then, in Sec.~III, we
give an overview of the $f(Q,T)$ gravity in the framework of
Friedmann-Lema\^{\i}tre-Robertson-Walker (FLRW). Sec.~IV is
devoted to the investigation of warm inflation within the context
of a linear version of $f(Q,T)$ gravity. In Sec.~V, we investigate
the cosmological perturbations originating from $f(Q,T)$ gravity
during warm inflation. Thereafter, in Sec.~VI, an analysis is
conducted regarding warm inflation featuring a power-law potential
in the strong dissipation regime, where constant and variable
dissipation coefficients are considered. At the end of this
section, we discuss the weak dissipation regime very briefly.
Finally, in Sec.~VII, we summarize the concluding results
obtained.

\section{Cosmological Warm Inflation}
 Let us first start with the action of GR in the presence of a
matter field as
\begin{equation}\label{inflation}
S=\int \sqrt{-g}\left(\dfrac{R}{2\kappa}+L_{\rm m}\right){\rm
d}^4x.
\end{equation}
Here, $g$ is the metric determinant, $R$ stands for the Ricci
scalar, $\kappa=1/M^2_{\rm Pl}=8\pi G$ where $M_{\rm Pl}$
signifies the reduced Planck mass in the natural units with
$\hbar=1=c$, and $L_{\rm m}$ denotes the Lagrangian of a matter
field. By varying action~\eqref{inflation} with respect to the
metric, one obtains
\begin{equation}\label{enesteineq}
R_{\mu \nu}-\dfrac{1}{2}R g_{\mu\nu}=\kappa\, T_{\mu\nu},
\end{equation}
where $T_{\mu\nu}$ represents the energy-momentum tensor generally
defined as
\begin{equation}\label{tmunu}
T_{\mu\nu}\equiv -\dfrac{2}{\sqrt{-g}}\frac{\delta(\sqrt{-g}L_{\rm
m})}{\delta g^{\mu\nu}}.
\end{equation}
We also consider the spatially flat FLRW geometry
\begin{equation}\label{frwmetric}
ds^{2}=-dt^{2}+a^{2}(t)(dx ^{2}+ dy ^{2}+dz^{2}),
\end{equation}
where $a(t)$ represents the cosmological scale factor at the
cosmic time $t$.

Assume that the matter is the inflaton scalar field with the
Lagrangian
\begin{equation}\label{inflatonlagrangy}
L_{\rm
m}^{[\phi]}=-\dfrac{1}{2}g^{\mu\nu}\partial_{\mu}\phi\,\partial_{\nu}\phi-V(\phi)=\dfrac{1}{2}\dot{\phi}^2-V(\phi),
\end{equation}
where we have used $\phi\equiv\phi(t)$ as a homogeneous scalar
field called inflaton, $V(\phi)$ is its potential, and the
lowercase Greek indices run from zero to three. Then, upon
inserting expression~\eqref{inflatonlagrangy} into
definition~\eqref{tmunu}, it gives
\begin{eqnarray}\label{Tphii}
T_{\mu\nu}^{[\phi]}&=&\partial_\mu\phi\,\partial_\nu\phi-g_{\mu\nu}\left[\dfrac{1}{2}
\partial_\sigma\phi\partial^\sigma\phi+V(\phi)\right]\cr
&=&\!\[\dfrac{1}{2}\dot{\phi}^2+V(\phi)\!\]\!\delta^0_\mu\delta^0_\nu+a^2\!\[\dfrac{1}{2}\dot{\phi}^2-V(\phi)
\!\]\!\delta^i_\mu\delta_{\nu i},
\end{eqnarray}
where the lowercase Latin indices run from one to three. The
energy-momentum tensor associated with the inflaton field (which
is conserved in this case) can be characterized as equivalent to a
perfect fluid with a linear barotropic equation of state
$p^{[\phi]}=w^{[\phi]}\rho^{[\phi]}$. Here, $\rho^{[\phi]}$ and
$p^{[\phi]}$ are, respectively, the energy density and the
pressure density of the inflaton field as
\begin{equation}\label{rho-p}
\rho^{[\phi]}=\dfrac{1}{2}\dot{\phi}^2+V(\phi)\qquad{\rm
and}\qquad p^{[\phi]}=\dfrac{1}{2}\dot{\phi}^2-V(\phi),
\end{equation}
where the dot represents the derivative with respect to the cosmic
time. Also, the trace of the energy-momentum tensor is
\begin{equation}\label{TraceT}
T^{[\phi]}=\dot{\phi}^2-4V.
\end{equation}
The variation of action~\eqref{inflation} with respect to the
inflaton field gives its corresponding field equation, and hence
utilizing all field equations, one obtains the corresponding
Friedmann equations.

Now, in the warm inflationary paradigm, the inflaton field
interacts with other fields during the inflationary
era~\cite{Berera1995,Berera2023}, and this leads to the
dissipation energy of inflaton as radiation. That is, the inflaton
field decays into radiation, and radiation is produced
continuously. Under these considerations, by employing
Eq.~\eqref{enesteineq} and relations \eqref{frwmetric},
\eqref{Tphii} and \eqref{rho-p}, while assuming that the radiation
field is a perfect fluid, we readily derive the new Friedmann
equations, which are amended in the case of warm inflation,
as~\cite{Berera1995}
\begin{equation}\label{firsth2}
3H^2=\kappa\(\rho^{[\phi]}+\rho^{[\gamma]}\),
\end{equation}
\begin{equation}\label{2ndFriedman}
2\dot{H}+3H^2=-\kappa\(p^{[\phi]}+p^{[\gamma]}\).
\end{equation}
Here, $H(t)\equiv\dot{a}(t)/a(t)$ is the Hubble parameter, and
$\rho^{[\gamma]}$ and $p^{[\gamma]}$ represent the energy density
and the pressure density associated with the radiation,
respectively. Actually, in the warm inflation, the background
source is
\begin{equation}\label{background}
\rho=\rho^{[\phi]}+\rho^{[\gamma]}\qquad{\rm and}\qquad
p=p^{[\phi]}+p^{[\gamma]},
\end{equation}
which is now conserved.

Analogous to damped harmonic oscillators, where the damping term
introduces dissipation effects and prevents energy conservation
with the rate of energy loss, it is common~\cite{BereraB2009,
Arya2018,BereraB2018} to consider the continuity equation
\begin{equation}\label{contp}
\dot{\rho}^{[\phi]}+3H\(\rho^{[\phi]}+p^{[\phi]}\)=-\Upsilon\dot{\phi}^2
\end{equation}
for the inflaton field in the warm inflation. Here, $\Upsilon$
represents the dissipation coefficient and, as a positive value in
accordance with the second law of thermodynamics, governs the
transition of energy from the inflaton into the radiation during
the inflationary era. The dissipation of inflaton into radiation
leads to the generation of entropy. With a thermodynamic potential
as $V(\phi,\tau)$, the thermodynamic relation defines the entropy
density of radiation, denoted by $s(\phi,\tau)$, in terms of the
thermodynamic potential as~\cite{Hall2004}
\begin{equation}\label{entropy}
s(\phi,\tau)=-\dfrac{\partial V(\phi,\tau)}{\partial\,\tau},
\end{equation}
where $\tau $ is the temperature of the thermal bath. Obviously,
if we consider the potential only as $V(\phi)$, the corresponding
entropy will vanish. Also, since the dissipation leads to the
generation of radiation, knowing that its pressure density is
$p^{[\gamma]}=\rho^{[\gamma]}/3$, its energy density changes over
time as
\begin{equation}\label{RadiatEqu}
\dot{\rho}^{[\gamma]}+4H\rho^{[\gamma]}=\Upsilon\dot{\phi}^2.
\end{equation}

Furthermore, by substituting relations~\eqref{rho-p} into
Eq.~\eqref{contp}, it gives the corresponding Klein-Gordon
equation as
\begin{equation}\label{kgeq}
\ddot{\phi}+\dot{\phi}\(3H+\Upsilon\)+V^{\prime}=0,
\end{equation}
where the prime denotes derivative with respect to the inflaton
field. Then, by defining the dissipation rate, which quantifies
the rate of the dissipation coefficient relative to the rate of
expansion as
\begin{equation}\label{dissipation rate}
\chi\equiv\dfrac{\Upsilon}{3H},
\end{equation}
we can rewrite Eq.~\eqref{kgeq} as
\begin{equation}\label{InflatEqu}
\ddot{\phi}+3H\dot{\phi}\(1+\chi\)+V^{\prime}=0.
\end{equation}

The $\chi\gg 1$ corresponds to strong dissipation regime, while
$\chi \ll 1$ corresponds to the domain of weak dissipation.
Additionally, various forms for the dissipation coefficient can be
considered, such as a constant $\Upsilon_0$, a function of the
inflaton field $\Upsilon(\phi)$, a function of the temperature of
the thermal bath $\Upsilon(\tau)$, and/or a combination of both
$\Upsilon(\phi,\tau)$~\cite{Berera1995, Berera1997, Zhang2009}. In
the following, we investigate the necessary conditions for warm
inflation.
 \newpage

\hskip0.7cm{\textbf{Slow-Roll Conditions and Parameters} }
 \vskip0.2cm

While considering relations \eqref{rho-p} and Eqs.
\eqref{firsth2}, \eqref{RadiatEqu} and \eqref{InflatEqu}, for warm
inflation to occur and last long enough, the conditions of
slow-roll must be satisfied as
\begin{align}
 \label{phiCondition}&\dot{\phi}^2\ll V(\phi),\qquad \ddot{\phi}\ll 3H\dot{\phi}(1+\chi)\\
\!\!\!\!\!\!\!\!\!\!\!\!\!\!\!\!\!\!\!\!\!\!\!\!\!\!\!\!\!\!\!\!\!\!\!\!\!\!\!\!\!\!\!\!\!\!\!\!\!\!\!\!\!\!\!\!\!\!\!\!\!\!\!%
\!\!\!\!\!\!\!\!\!\!\!\!\!\!\!\!\!\!\!\!\!\!\!\!\!\!\!\!\!\!\!\!\!\!\!\!\!\!\!\!\!\!\!\!\!\!\!\!\!\!\!\!\!{\rm and}\nonumber\\
 \label{radCondition}&\rho^{[\gamma]}\ll\rho^{[\phi]},\qquad
\dot{\rho}^{[\gamma]}\ll 4H\rho^{[\gamma]}\ \, {\rm and}\ \,
\Upsilon\dot{\phi}^2.
\end{align}
Here, the last two terms indicate the dominance of the inflaton
energy density over the radiation energy density during the warm
inflationary epoch and also imply the state of relative stability
in radiation generation~\cite{Berera1995, Berera2000, Hall2004}.

Under these considerations, Eqs. \eqref{firsth2},
\eqref{InflatEqu} and \eqref{RadiatEqu} are respectively reduced
to
\begin{align}
\label{fh2}& 3H^2\approx\kappa\, V,\\
\label{appkg}&3H\dot{\phi}(1+\chi)\approx -V^\prime, \\
\label{bbradiation}&\rho^{[\gamma]}\approx\dfrac{\Upsilon\dot{\phi}^2}{4H}=\dfrac{3}{4}\chi\dot{\phi}^2\mathrel{\mathop=^{\rm
also}}C_{\gamma}\tau^4,
\end{align}
where $C_{\gamma} = g_{\ast} \pi^2/30$ represents the
Stefan-Boltzmann constant for the black-body
radiation~\cite{BereraB2009}. Also, the constant $g_{\ast}$
denotes the effective number of relativistic degrees of freedom of
the radiation energy density in the thermal bath at the
temperature $\tau $, whose value is approximately $106.75$ for the
standard model particles~\cite{Correa2022}. In addition, relations
\eqref{rho-p} read
\begin{equation}\label{rho-pApproxi}
\rho^{[\phi]}\approx V(\phi)\approx
-p^{[\phi]}\quad\Rightarrow\quad \rho^{[\phi]}+p^{[\phi]}\approx 0
\end{equation}
during inflation. Also, by differentiating Eq.~\eqref{fh2} and
employing the Klein-Gordon Eq. \eqref{appkg}, we obtain the
equation for $\dot{H}$ as
\begin{equation}\label{firsthdot}
\dot{H}\approx -\kappa\(1+\chi\)\dfrac{\dot{\phi}^2}{2}.
\end{equation}

Following Refs.~\cite{liddle1994,Berera2009,
Visinelli2011,Setare2013,myrzak}, there are several parameters
that can be used to generally parameterize the slow-roll
approximations, namely
\begin{align}
&\label{eps1}\epsilon\equiv-\dfrac{\dot{H}}{H^2},\\
&\label{et1} \eta\equiv-\dfrac{\dot{H}}{H^2}-\dfrac{\ddot{H}}{2\dot{H}H},\\
&\label{lam1}\lambda\equiv-\dfrac{\dot{\rho}^{[\gamma]}}{H\rho^{[\gamma]}}.
\end{align}
Warm inflation occurs when these slow-roll parameters meet the
conditions
\begin{equation}\label{conditions}
\epsilon\ll 1,\qquad \mid\eta\mid \ll 1, \qquad \mid\lambda\mid\ll 1.
\end{equation}
However, when the inflation ends, the value of the slow-roll
parameters becomes one. Utilizing Eqs. \eqref{fh2}, \eqref{appkg},
\eqref{bbradiation} and \eqref{firsthdot} plus conditions
\eqref{conditions}, we can rephrase the slow-roll parameters in
terms of the potential at leading order as
\begin{align}
\label{eps}&\epsilon\approx\dfrac{1}{2\kappa\(1+\chi\)}\(\dfrac{V^\prime}{V}\)^2\equiv\dfrac{\epsilon_{V}}{1+\chi}, \\
\label{et}&\eta\approx\dfrac{1}{\kappa\(1+\chi\)}\(\dfrac{V^{\prime\prime}}{V}\)\equiv\dfrac{\eta_{V}}{1+\chi},\\
\label{lam}&\lambda\approx\dfrac{1}{\kappa\(1+\chi\)}\(\dfrac{\Upsilon^\prime
V^\prime}{\Upsilon V}\)\equiv\dfrac{\lambda_{V}}{1+\chi},
\end{align}
where obviously $1+\chi > 0$ because $\chi $ is positive by
definition, and we have assumed that $\Upsilon $ is, in general,
an explicit function of the inflaton field. Therefore, for warm
inflation to occur, the conditions $\epsilon_{V}\ll 1+\chi$,
$\eta_{V}\ll 1+\chi$ and $\lambda_{V}\ll 1+\chi$ must hold.

Moreover, using Eqs.~\eqref{fh2} and \eqref{appkg}, we can write
the radiation energy density in terms of the potential as
\begin{equation}
\rho^{[\rm
\gamma]}\approx\dfrac{\chi}{4\kappa\(1+\chi\)^2}\dfrac{\(V^\prime\)^2}{V}.
\end{equation}
Subsequently, by employing, respectively, relations \eqref{eps},
\eqref{phiCondition} and \eqref{rho-p}, we obtain the relation
between the radiation energy density and the inflaton energy
density during warm inflation as
\begin{equation}\label{rad1}
\rho^{[\rm \gamma]}\approx\dfrac{\chi}{2\(1+\chi\)}\epsilon V\approx \dfrac{\chi}{2\(1+\chi\)}\epsilon \rho^{[\phi]}.
\end{equation}
This relation is self-consistent with the dominance of the
inflaton energy density over the radiation energy density during
the warm inflationary epoch, relation \eqref{radCondition}. At the
end of inflation, in the strong dissipative regime, $\rho^{[\rm
\gamma]} \approx \rho^{[\phi]}/2$, and in the weak dissipative
regime, still $\rho^{[\gamma]}\ll\rho^{[\phi]}$.

Another significant parameter during inflation is the e-folding
number. This parameter quantifies the extent of expansion of the
Universe throughout the inflationary epoch and is generally
defined as~\cite{Liddle1999, Baumann2007}
\begin{equation}\label{Ndef}
N\equiv \ln\(\dfrac{a_{\rm end}}{a}\)=\int_{t}^{t_{\rm end}}
H\rm{dt}.
\end{equation}
Here, the subscript `${\rm end}$' denotes the value of the
quantities at the end of inflation. Utilizing Eqs.~\eqref{fh2} and
\eqref{appkg}, the e-folding number in the case of warm inflation
can be reformulated in terms of the potential as
\begin{equation}
N=\int_{\phi}^{\phi_{\rm end}}
\dfrac{H}{\dot{\phi}}\rm{d\phi}\approx\kappa \int_{\phi_{\rm
end}}^{\phi}\(1+\chi\)\dfrac{V}{V^\prime}\rm{d\phi}.
\end{equation}

In the subsequent sections, we first provide a concise overview of
the theoretical underpinnings of $f(Q, T)$ gravity\footnote{For
more details, see, e.g., Refs.~\cite{Xu2019,Shiravand2022} and
references therein.}\
 and then investigate its cosmological consequences.

\section{$f(Q, T)$ Gravity: a brief review}
At first, it is known that any general connection can be
decomposed into three independent components, namely
\begin{equation}\label{affine}
\Gamma^{\alpha}{}_{\mu\nu}=\{^{\alpha}{}_{\mu\nu}\}+K^{\alpha}{}_{\mu\nu}+L^{\alpha}{}_{\mu\nu},
\end{equation}
where $\{^{\alpha}{}_{\mu\nu}\}$, $K^{\alpha}{}_{\mu\nu}$ and
$L^{\alpha}{}_{\mu\nu}$ represent the Christoffel symbol, the
contorsion tensor, and the disformation tensor, respectively.
These terms are
\begin{eqnarray}
\label{leci}\{^{\alpha}{}_{\mu\nu}\}&=&\dfrac{1}{2}g^{\alpha\beta}(\partial_{\mu}
g_{\beta\nu} +\partial _{\nu}g_{\beta\mu}-\partial_{\beta}
g_{\mu\nu}),\cr
K^{\alpha}{}_{\mu\nu}&=&\dfrac{1}{2}\mathbb{T}^{\alpha}{}_{\mu\nu}+\mathbb{T}_{(\mu\nu)}{}^{\alpha},\cr
L^{\alpha}{}_{\mu\nu}&\equiv
&-\dfrac{1}{2}g^{\alpha\beta}\left(Q_{\mu\beta\nu}
+Q_{\nu\beta\mu}-Q_{\beta\mu\nu}\right),
\end{eqnarray}
where $\mathbb{T}^{\alpha}{}_{\mu\nu}=
2\Gamma^{\alpha}{}_{[\mu\nu]}$ represents the torsion tensor and
$Q_{\alpha\mu\nu}\equiv\nabla_{\alpha} g_{\mu\nu}\neq 0$ stands
for the nonmetricity tensor.

Then, for later use, we define the traces of nonmetricity tensor,
the superpotential tensor and the nonmetricity scalar,
respectively, as
\begin{equation}
{Q}_{\alpha}\equiv Q_{\alpha}{}^{\mu}{}_{\mu}\qquad{\rm
and}\qquad\tilde{Q}_{\alpha}\equiv Q^{\mu}{}_{\alpha\mu},
\end{equation}
\begin{equation}
P^{\alpha}{}_{\mu\nu}\equiv
-\dfrac{1}{2}L^{\alpha}{}_{\mu\nu}+\dfrac{1}{4}
\left(Q^{\alpha}-\tilde{Q}^{\alpha}\right)g_{\mu\nu}-\dfrac{1}{4}\delta^{\alpha}_{(\mu}Q_{\nu)},
\end{equation}
\begin{equation}\label{invariant}
Q\equiv -g^{\mu
\nu}\left(L^{\alpha}{}_{\beta\mu}L^{\beta}{}_{\nu\alpha}-L^{\alpha}{}_{\beta\alpha}L^{\beta}{}_{\mu\nu}\right).
\end{equation}
However, in the case of vanishing the Riemann and torsion tensors,
i.e. a flat manifold case, there is always an adapted coordinate
system (i.e., the coincident gauge) where $\nabla_{\mu}
\stackrel{*}{\rightarrow}
\partial_{\mu}$, that is the connection effectively vanishes.
Consequently, in this particular system, relations \eqref{affine}
and \eqref{invariant} read
\begin{equation}
L^{\alpha}{}_{\mu\nu} \mathrel{\mathop=^{*}}
-\{^{\alpha}{}_{\mu\nu}\},
\end{equation}
\begin{equation}\label{Qstar}
Q \mathrel{\mathop=^{*}}
-g^{\mu\nu}\Big(\{^{\alpha}{}_{\beta\mu}\}\{^{\beta}{}_{\nu\alpha}\}-
\{^{\alpha}{}_{\beta\alpha}\}\{^{\beta}{}_{\mu\nu}\}\Big),
\end{equation}
where the second result is equal to the minus of the effective
Einstein-Hilbert Lagrangian term. Moreover, relation \eqref{Qstar}
can be easily derived with metric \eqref{frwmetric} and obtain
\begin{equation}\label{QasH}
Q=6H^2,
\end{equation}
which is valid as a scalar in any frame.

Now, let us investigate the action for $f(Q,T)$ gravity in the
presence of the inflaton field as
\begin{equation}\label{fqtaction}
S=\int \sqrt{-g}\left[\dfrac{f(Q, T)}{2\kappa}+L_{\rm
m}^{[\phi]}\right]{\rm d}^{4}x,
\end{equation}
where its dynamical variables are the inflaton field, the metric,
and the connection, which we assume to be in the case of vanishing
the Riemann and torsion tensors. Under these assumptions, varying
action \eqref{fqtaction} with respect to the metric and the
connection, respectively, gives the field equations
\begin{equation}\label{fieldeq}
\begin{array}{l}
\kappa\,
T_{\mu\nu}^{[\phi]}=-\dfrac{2}{\sqrt{-g}}\bigtriangledown_{\alpha}
\left(f_{\rm Q}\sqrt{-g}P^{\alpha}{}_{\mu\nu}\right)-\dfrac{f\,g_{\mu\nu}}{2}\\ \\
\kern 1.4pc +f_{\rm
T}\left(T_{\mu\nu}^{[\phi]}+\Theta_{\mu\nu}\right) -f_{\rm
Q}\left(P_{\mu\alpha\beta} Q_{\nu}{}^{\alpha\beta}-2\,
Q^{\alpha\beta}{}_{\mu}P_{\alpha\beta\nu}\right)
\end{array}
\end{equation}
and
\begin{equation}\label{anotherFieldEq}
\bigtriangledown_{\mu}\bigtriangledown_{\nu}\(2\sqrt{-g}f_{\rm
Q}P^{\mu\nu}{}_{\alpha}+\kappa\, {\cal H}_{\alpha}{}^{\mu\nu}\)=0,
\end{equation}
where
\begin{equation}
\Theta_{\mu\nu}\equiv g^{\alpha\beta}\dfrac{\delta
T_{\alpha\beta}^{[\phi]}} {\delta g^{\mu\nu}},~~ f_{\rm
Q}\equiv\dfrac{\partial f(Q,T)}{\partial Q},~~ f_{\rm
T}\equiv\dfrac{\partial f(Q,T)}{\partial T}
\end{equation}
and
\begin{equation}
{\cal H}_{\alpha}{}^{\mu\nu}\equiv\dfrac{f_{\rm
T}\sqrt{-g}}{2\kappa}\dfrac{\delta\,
T^{[\phi]}}{\delta\Gamma^{\alpha}{}_{\mu\nu}}+\dfrac{\delta\(\sqrt{-g}\,
L_{\rm m}^{[\phi]}\)}{\delta \Gamma^{\alpha}{}_{\mu\nu}}.
\end{equation}

Considering all the assumptions and after some manipulations,
these field equations lead to the modified Friedmann equations
\begin{align}
\label{rho}&\kappa\,\rho^{[\phi]}=\dfrac{f}{2}-6FH^2-\dfrac{2\tilde{G}}{1+\tilde{G}}\left(\dot{F}H+F\dot{H}\right),\\
\label{p}&\kappa\,
p^{[\phi]}=-\dfrac{f}{2}+6FH^2+2\left(\dot{F}H+F\dot{H}\right),
\end{align}
where $F\equiv f_{\rm Q}$ and $\kappa\tilde{G}\equiv f_{\rm T}$.
By introducing the effective energy and pressure densities, as a
conserved perfect fluid, we rewrite the above equations as
\begin{align}\label{rhoeff}
&3H^2=\dfrac{f}{4F}-\dfrac{\kappa}{2F}\left[\left(1+\tilde{G}\right)\rho^{[\phi]}
+\tilde{G}p^{[\phi]}\right]\equiv\kappa\rho^{\rm [eff]},\\
&2\dot{H}+3H^2=\dfrac{\kappa}{2F}\left[\left(1+\tilde{G}\right)\rho^{[\phi]}+\left(2+\tilde{G}\right)p^{[\phi]}\right]\nonumber\\
\label{peff}&~~~~~~~~~~~~~~~~~+\dfrac{f}{4F}-\dfrac{2\dot{F}H}{F}\equiv
-\kappa p^{\rm [eff]},
\end{align}
Also, the evolution of the Hubble parameter is
\begin{equation} \label{Hdot}
\dot{H}+\frac{\dot{F}}{F}H=\frac{\kappa}{2F}\left(1+\tilde{G}\right)
\left(\rho^{[\phi]}+p^{[\phi]}\right).
\end{equation}

Furthermore, we recall that the energy-momentum tensor of the
source, $T_{\mu\nu}^{[\phi]}$, is~not conserved in the context of
the $f(Q,T)$ gravity. To illustrate this, let us first rewrite the
field equation \eqref{fieldeq} as
\begin{align}
&f_{\rm T} \left(T^{[\phi]\mu}{}_{\nu}+\Theta^{\mu}{}_{\nu}\right)-\kappa T^{[\phi]\mu}{}_{\nu}\nonumber\\
&=\dfrac{f}{2}\delta^{\mu}{}_{\nu}+f_{\rm Q}
Q_{\nu}{}^{\alpha\beta} P^{\mu}{}_{\alpha\beta}
+\dfrac{2}{\sqrt{-g}}\nabla_\alpha\left(f_{\rm Q} \sqrt{-g}
P^{\alpha\mu}{}_{\nu}\right).
\end{align}
Then, while using the field equation \eqref{anotherFieldEq}, the
covariant divergence of $T_{\mu\nu}^{[\phi]}$ can be obtained from
the above equation, which becomes~\cite{Antonio2022,Xu2019}
\begin{align}\label{covleviT}
&\mathcal{D}_{\mu} T^{[\phi]\mu}{}_{\nu}-\dfrac{1}{f_{\rm
T}+\kappa}\left(f_{\rm T} \partial_\nu p^{[\phi]}
-\dfrac{1}{2}f_{\rm T}
\partial_\nu T^{[\phi]}-B_\nu \right)=0,
\end{align}
where $f_{\rm T}+\kappa\neq 0$, the symbol $\mathcal{D}_{\mu}$
signifies the covariant derivative with respect to the Levi-Civita
connection, and
\begin{align}
B_{\nu}\equiv &\dfrac{1}{\sqrt{-g}}\Big[Q_\mu \nabla_\alpha \Big(\sqrt{-g}f_{\rm Q} P^{\alpha\mu}{}_{\nu}\Big)\nonumber\\
&+2 \nabla_\mu \nabla_\alpha\Big(\sqrt{-g}f_{\rm Q}
P^{\alpha\mu}{}_{\nu}\Big)\Big].
\end{align}

Eq. \eqref{covleviT} also indicates that in the case of metricity,
i.e. when $B_{\nu}=0$, plus lack of the matter-curvature coupling,
i.e. when $f_T=0$, the energy-momentum tensor of the source is
conserved. However, finding a $T^{\rm [eff]}_{\mu\nu}$ to write
Eq. \eqref{covleviT} in the form of its covariant divergence equal
to zero is~not an easy task. That is why we resort to the
definition of $\rho^{\rm [eff]}$ and $p^{\rm [eff]}$, as given in
Eqs. \eqref{rhoeff} and \eqref{peff}, to define such an effective
conserved term.

\section{Warm Inflation in $f(Q,T)$ Gravity}
In this section, we investigate the warm inflationary concept
within the framework of the linear version of $f(Q,T)$ gravity in
the presence of the inflaton scalar field $\phi$. In fact, we
consider the specific functional form of $f(Q,T)$ gravity as
\begin{equation}\label{linearf}
f(Q,T)=\alpha\, Q+\beta\, T^{[\phi]},
\end{equation}
where $\alpha$ and $\beta$ are constants within the model.

Analogous to Sec.~II, the dynamics of the warm inflation scenario
in the framework of $f(Q,T)$ gravity can be described as
\begin{align}
\label{WarmFried1}&3H^2=\kappa\left(\rho^{[\rm eff]}+\rho^{[\gamma]}\right),\\
\label{WarmFried2}&2\dot{H}+3H^2=-\kappa \left(p^{[\rm
eff]}+p^{[\gamma]}\right).
\end{align}
Although the inflaton field decays into radiation, due to Eqs.
\eqref{rhoeff} and \eqref{peff}, we have
\begin{align}
\label{conteff}&\dot{\rho}^{[\rm eff]}+3H\(\rho^{[\rm
eff]}+p^{[\rm eff]}\)
=-\Upsilon\dot{\phi}^2,\\
\label{contRadi}&\dot{\rho}^{[\rm \gamma]}+4H\rho^{[\rm
\gamma]}=\Upsilon\dot{\phi}^2.
\end{align}
Furthermore, while considering the slow-roll condition $\rho^{[\rm
\gamma]}\ll \rho^{[\rm \phi]}$, upon utilizing
relations~\eqref{rho-p}, \eqref{TraceT}, \eqref{QasH},
\eqref{Hdot} and \eqref{linearf} into Eqs.~\eqref{WarmFried1} and
\eqref{WarmFried2}, we obtain
\begin{align}
\label{rho11}&3H^2\approx
-\dfrac{\left(\kappa+\beta\right)\dot{\phi}^2+
2V(\phi)\left(\kappa+2\beta\right)}{2\alpha}\approx\kappa\rho^{[\rm eff]},\\
\label{p11}&2\dot{H}+3H^2\approx\dfrac{\left(\kappa+\beta\right)
\dot{\phi}^2-2V(\phi)\left(\kappa+2\beta\right)}{2\alpha}\approx
-\kappa p^{[\rm eff]}.
\end{align}
Then, by substituting Eqs. \eqref{rho11} and \eqref{p11} into
Eq.~\eqref{conteff}, we can derive the corresponding modified
Klein-Gordon equation
\begin{equation}\label{kgp}
\ddot{\phi}\left(\kappa+\beta\right)+3H\dot{\phi}\left(\kappa+\beta-\kappa\alpha\chi\right)
+V^{\prime}\left(\kappa+2\beta\right)\approx 0.
\end{equation}
In addition, by differentiating Eq. \eqref{rho11} while using Eq.
\eqref{kgp}, we obtain the modified equation governing $\dot{H}$
as
\begin{equation}\label{HHdot}
\dot{H}\approx
-\(\frac{\kappa+\beta-\kappa\alpha\chi}{-\alpha}\)\dfrac{\dot{\phi}^2}{2}
\end{equation}
that modifies Eq. \eqref{Hdot} in this case for the warm inflation
scenario. Actually, the last term on the right-hand side of Eq.
\eqref{HHdot} comes from Eq. \eqref{bbradiation}.

Also, in this model, the slow-roll conditions \eqref{phiCondition}
should be modified as
\begin{align}
&\(\kappa+\beta\)\dot{\phi}^2\ll 2V\(\kappa+2\beta\),\\
&\(\kappa+\beta\)\ddot{\phi}\ll
3H\dot{\phi}\(\kappa+\beta-\kappa\alpha\chi\).
\end{align}
By employing these approximations, Eqs. \eqref{rho11} and
\eqref{kgp} read
\begin{align}
\label{h2ap}&3H^2\approx \dfrac{-\(\kappa+2\beta\)}{\alpha}V,\\
\label{appk}&3H\dot{\phi}\(\kappa+\beta-\kappa\alpha\chi\)\approx
-\(\kappa+2\beta\)V^\prime.
\end{align}
Then, using Eqs.~\eqref{bbradiation}, \eqref{HHdot}, \eqref{h2ap}
and \eqref{appk} plus conditions \eqref{conditions} into relations
\eqref{eps1}, \eqref{et1} and \eqref{lam1}, we can derive the
modified slow-roll parameters within the linear version of
$f(Q,T)$ gravity~\eqref{linearf} in terms of the inflaton
potential as
\begin{align}
\label{epsv}&\epsilon\approx\dfrac{-\alpha}{2\(\kappa+\beta-\kappa\alpha\chi\)}\(\dfrac{V^\prime}{V}\)^2,\\
\label{etv}&\eta\approx\dfrac{-\alpha}{\kappa+\beta-\kappa\alpha\chi}\(\dfrac{V^{\prime\prime}}{V}\),\\
\label{lamv}&\lambda\approx\dfrac{-\alpha}{\kappa+\beta-\kappa\alpha\chi}\(\dfrac{\Upsilon^\prime
V^\prime}{\Upsilon V}\),
\end{align}
where $\kappa+\beta-\kappa\alpha\chi \neq 0$. As is obvious, the
comparison of these parameters with the corresponding
relations~\eqref{eps}, \eqref{et}, and \eqref{lam} shows that the
slow-roll parameters have been modified by
\begin{equation}\label{constraint}
\dfrac{1}{\kappa\(1+\chi\)}\longrightarrow\dfrac{-\alpha}{\kappa+\beta-\kappa\alpha\chi}.
\end{equation}
Moreover, the obtained results dictate the constraint
\begin{equation}\label{constrainteq}
\dfrac{-\alpha}{\kappa+\beta-\kappa\alpha\chi} >0,
\end{equation}
which, in turn, for negative values of $\alpha $, we must have
\begin{equation}\label{NEWconstrainteq}
\kappa+\beta-\kappa\alpha\chi >0.
\end{equation}

Additionally, by using Eqs. \eqref{h2ap}, \eqref{appk} and
\eqref{epsv} into Eq. \eqref{bbradiation}, we obtain the radiation
energy density in this case as
\begin{equation}\label{RhoRadAgain}
\rho^{[\rm \gamma]}\approx
\dfrac{\chi\(\kappa+2\beta\)}{2\(\kappa+\beta-\kappa\alpha\chi\)}
\epsilon V\approx
\dfrac{-\kappa\alpha\chi}{2\(\kappa+\beta-\kappa\alpha\chi\)}\epsilon\rho^{[\rm
eff]},
\end{equation}
where the last approximation is done employing Eq. \eqref{h2ap}
with approximation \eqref{rho11}. Relation \eqref{RhoRadAgain}
shows that $\rho^{[\rm \gamma]}\ll \rho^{[\rm eff]}$, which is
consistent with the condition $\rho^{[\rm \gamma]}\ll \rho^{[\rm
\phi]}$. Also, at the end of inflation, in the strong dissipative
regime, $\rho^{[\rm \gamma]}\approx \rho^{[\rm eff]}/2$, and in
the weak dissipative regime, still $\rho^{[\rm \gamma]}\ll
\rho^{[\rm eff]}$. In addition, utilizing Eqs.~\eqref{h2ap} and
\eqref{appk}, the e-folding number in this case is
\begin{equation}\label{nv}
N\approx\dfrac{1}{-\alpha}\int_{\phi_{\rm
end}}^{\phi}\(\kappa+\beta-\kappa\alpha\chi\)\dfrac{V}{V^{\prime}}d\phi.
\end{equation}

\section{COSMOLOGICAL PERTURBATIONS}
The observations acquired from the CMB show that the power
spectrum of primordial perturbations aligns well with the power
spectrum predictions in the inflationary theory. In the cold
inflation model, these perturbations stem from the quantum
fluctuations of the inflaton field. However, in the context of
warm inflation, its significance lies in the involvement of
dissipative effects throughout the inflationary period. This
action leads to the simultaneous occurrence of radiation
production and inflationary expansion. Dissipation is attributed
to a friction term that characterizes the process of dissipating a
scalar field into a thermal bath through its interaction with
other fields. Moreover, warm inflation illustrates the influential
role of thermal fluctuations in generating initial perturbations
during inflation. In these models, density fluctuations stem
primarily from thermal fluctuations rather than quantum
fluctuations, and these fluctuations constitute the initial
conditions for the formation of large-scale
structures~\cite{Berera1999, Berera2009}. In the following
subsections, we first prepare the necessary equations for scalar
and tensor perturbations in the context of warm inflation within
the $f(Q,T)$ gravity. Then, we probe the effects of these
perturbations on the linear version of $f(Q,T)$ gravity
relation~\eqref{linearf}.

\subsection{SCALAR PERTURBATIONS}
In this subsection, we explore an equation that governs
cosmological perturbations within a system that encompasses both
the inflaton field and radiation. In the process of perturbation,
we mainly perturb four quantities, namely the metric, the
four-velocity of the radiation fluid, the inflaton field and the
energy density of the radiation field. Of course, these
perturbations in turn cause the other quantities involved to be
perturbed. Furthermore, our particular focus is up to the
second-order of perturbations; and, to be consistent with the
literature, from now on, we show unperturbed quantities with a bar
over them.

For perturbing the metric in the Einstein frame by adopting the
Newtonian/longitudinal gauge, the perturbation of the background
(unperturbed) metric~\eqref{frwmetric}, when the energy-momentum
tensors involved are diagonal, is expressed as~\cite{Hall2004,
Mukhanov1992, Bari2019}
\begin{equation}\label{30}
ds^2=-(1+2\Psi)dt^2+a^2(t)(1-2\Psi)\delta_{ij}dx^idx^j.
\end{equation}
Here, the metric perturbation variable is assumed to be
$|\Psi(t,{\bf x})|\ll 1$, which parameterizes the scalar
perturbations. For later use, from the perturbed
metric~\eqref{30}, we obtain
\begin{equation}\label{upMetricPerturb}
\delta g^{\mu\nu}\approx {\rm
diag}\left[2\Psi,2\Psi/a^2,2\Psi/a^2,2\Psi/a^2\right].
\end{equation}

In the warm inflation scenario, it is possible to divide the
energy-momentum tensor into two components, that is, the effective
component (which consists of the inflaton part and the correction
terms related to the $f(Q,T)$ theory, i.e. Eqs.~\eqref{rhoeff} and
\eqref{peff}) and the radiation part denoted by
$T_{\mu\nu}^{[\gamma]}$. That is, we have
\begin{equation}\label{31}
T_{\mu\nu}=T_{\mu\nu}^{[\gamma]}+T_{\mu\nu}^{[\rm eff]}.
\end{equation}
As mentioned, the radiation field is described as a perfect
barotropic fluid, i.e.
\begin{equation}\label{32}
T_{\mu\nu}^{[\gamma]}=\frac{\rho^{[\gamma]}}{3}\(4u_{\mu}u_{\nu}+g_{\mu\nu}\),
\end{equation}
with the same relation for the corresponding unperturbed one. In
this regard, the components of the unperturbed four-velocity of
the radiation fluid, $\overline{u}_{\mu}$, are $\overline{u}_0=-1$
and $\overline{u}_i=0$ in the rest frame. For perturbing the
four-velocity of the radiation fluid, we assume that the perturbed
four-velocity of the radiation fluid is
\begin{equation}\label{33}
u_{\mu}(t,{\bf x})=\overline{u}_{\mu}(t)+\delta u_{\mu}(t,{\bf x})
\end{equation}
with the normalization constraint
\begin{equation}\label{normalization}
u^{\alpha}u_{\alpha}=g^{\mu\nu}u_{\mu}u_{\nu}=-1,
\end{equation}
where, obviously, $u^{\mu}\approx \overline{u}^{\mu}+\delta
u^{\mu}+\overline{u}_\nu\delta g^{\mu\nu}$. The variable $\delta
u_{\mu}$ represents the perturbed part of the four-velocity of the
radiation fluid, where $\delta
u^\mu\equiv\delta(u^\mu)=\delta(g^{\mu\nu}u_\nu)$~\cite{Weinberg},
and constraint~\eqref{normalization} gives
\begin{equation}\label{34}
\delta u_0 \approx -\Psi\qquad \Rightarrow \qquad \delta u^0
\approx -\Psi .
\end{equation}
However, the velocity perturbation, $\delta u_i$, stands as an
independent dynamic variable that can\footnote{This is possible
because for scalar perturbations, such a `flow' is irrotational.}\
 be defined as
\begin{equation}\label{delta_u}
\delta u_i \equiv a(t)\partial_i(\delta {\mathfrak u})\qquad
\Rightarrow \qquad \delta u^i \approx a(t)\partial^i(\delta
{\mathfrak u}),
\end{equation}
where $\delta {\mathfrak u} $ is a scalar velocity
potential~\cite{Weinberg,mukhanov}.

For perturbing the inflaton field and the energy density of the
radiation field, we assume that these perturbed fields are
\begin{equation}\label{perturbPhi}
\phi(t,{\bf x})=\overline{\phi}(t)+\delta \phi(t,{\bf x}),
\end{equation}
\begin{equation}\label{perturbRadiation}
\rho^{[\gamma]}(t,{\bf x})=\overline{\rho^{[\gamma]}}(t)+\delta
\rho^{[\gamma]}(t,{\bf x}),
\end{equation}
and in turn, we have $\delta p^{[\gamma]}=\delta
\rho^{[\gamma]}/3$.

Next, the exchange of energy between the two components
$\rho^{[\phi]}$ (and as a result $\rho^{[\rm eff]}$) and
$\rho^{[\gamma]}$ can be depicted through a flux term
as~\cite{Moss2007}
\begin{equation}\label{35}
\Lambda_{\mu}=-\Upsilon u^{\nu}\partial_{\mu}\phi\,
\partial_{\nu}\phi.
\end{equation}
From the same relation for its corresponding unperturbed one, we
obtain
$\overline{\Lambda}_{0}=-(\dot{\overline{\phi}})^2\,\overline{\Upsilon}$
and $\overline{\Lambda}_{i}=0$. Hence, Eqs. \eqref{conteff} and
\eqref{contRadi} read
\begin{align}
\label{conteff2}&\dot{\overline{\rho^{[\rm
eff]}}}+3H\(\overline{\rho^{[\rm eff]}}+\overline{p^{[\rm eff]}}\)
= -\overline{\Lambda}^{0},\\
\label{contRadi2}&\dot{\overline{\rho^{[\rm
\gamma]}}}+4H\overline{\rho^{[\rm
\gamma]}}=\overline{\Lambda}^{0}.
\end{align}
In the same merit, and in general, the conservation of the
corresponding energy-momentum tensors can be written as
\begin{eqnarray}
\label{75}\overline{{\cal D}}_{\mu}\overline{T^{[\rm eff]}}^{\mu}{}_{\nu}& = &-\overline{\Lambda}_{\nu},\\
\label{36} \overline{{\cal
D}}_{\mu}\overline{T^{[\gamma]}}^{\mu}{}_{\nu}& =
&\overline{\Lambda}_{\nu},
\end{eqnarray}
with the same equations for the corresponding perturbed ones.

At this stage, we prepare some necessary equations for use in the
next steps. By writing the perturbed exchange of energy as
$\Lambda_\mu=\overline{\Lambda}_\mu+\delta\Lambda_\mu$ and
substituting all terms into relation~\eqref{35}, we obtain the
components of perturbation of exchange energy as\footnote{Although
relation $[\partial_\mu,\delta]=0$ holds, we do~not use it to
avoid mixing up functions, e.g., $\phi$ with $\delta\phi$, where
the former actually is $\phi(t)$ and the latter $\delta\phi(t,{\bf
x})$.}
\begin{eqnarray}\label{37}
\delta \Lambda_{0}&=&-
(\dot{\overline{\phi}})^2\,\delta\Upsilon+(\dot{\overline{\phi}})^2\,\Psi\overline{\Upsilon}
-2\overline{\Upsilon}\,\dot{\overline{\phi}}\,\partial_0(\delta\phi),\\
\label{78}\delta
\Lambda_{i}&=&-\overline{\Upsilon}\,\dot{\overline{\phi}}\,\partial_i(\delta\phi).
\end{eqnarray}
Also, by perturbing Eq. \eqref{36} for the corresponding perturbed
quantities, i.e. $\delta\({\cal
D}_{\mu}T^{[\gamma]\mu}{}_{\nu}\)=\delta \Lambda_{\nu}$, while
using metric \eqref{30} (to compute the Levi-Civita connection),
relations \eqref{32}, \eqref{33} and \eqref{perturbRadiation}, and
Eqs. \eqref{contRadi2}, \eqref{37} and \eqref{78}, we obtain from
its zeroth and i-th components, respectively
\begin{align}\label{38}
&\partial_0(\delta\rho^{[\gamma]})+4H\delta\rho^{[\gamma]}+{4\over3}\overline{\rho^{[\gamma]}}a\,\nabla^2(\delta
{\mathfrak u})
-4\dot{\Psi}\overline{\rho^{[\gamma]}}\nonumber\\
&\approx \delta\Upsilon
(\dot{\overline{\phi}})^2-\Psi\overline{\Upsilon}\,(\dot{\overline{\phi}})^2
+2\overline{\Upsilon}\,\dot{\overline{\phi}}\,\partial_0(\delta\phi)
\end{align}
and
\begin{align}\label{38.2}
&4\left[\overline{\rho^{[\gamma]}}\partial_0\partial_i(\delta
{\mathfrak u})+{\dot{\overline{\rho^{[\gamma]}}}}\partial_i(\delta
{\mathfrak u})+4\, H\overline{\rho^{[\gamma]}}\partial_i(\delta
{\mathfrak u})\right]\nonumber\\
&\approx
-a^{-1}\left[3\overline{\Upsilon}\,\dot{\overline{\phi}}\partial_i(\delta\phi)+\partial_i(\delta\rho^{[\gamma]})
+4\,\overline{\rho^{[\gamma]}}\partial_i\Psi\right],
\end{align}
where $\nabla^2=\partial_i\partial^i$.

Furthermore, by perturbing the field equation \eqref{fieldeq}
while using all necessary equations, one can
obtain~\cite{Antonio2022} from its $0-0$ component, trace, and
$0-i$ components, respectively
\begin{align}\label{44}
&\delta\rho^{[\phi]}\Big[\kappa+{3\over2}\overline{f_{\rm
T}}-\overline{f_{\rm TT}}\Big(\overline{\rho^{[\phi]}}
+\overline{p^{[\phi]}}\Big)\Big]\!\nonumber\\
&+\delta p^{[\phi]}\Big[\!-{1\over2}\overline{f_{\rm T}}+3\overline{f_{\rm TT}}\(\overline{\rho^{[\phi]}}
+\overline{p^{[\phi]}}\)\!\Big]\nonumber\\
\approx &
\, 2\overline{f_{\rm Q}}\(3H^2\Psi+3H\dot{\Psi}-a^{-2}\nabla^2\Psi\)\nonumber\\
&+72H^3\overline{f_{\rm QQ}}\(H\Psi+\dot{\Psi}\),
\end{align}
\begin{align}\label{45}
&{\dfrac{1}{4}}\overline{f_{\rm
T}}\delta\rho^{[\phi]}-\Big(\frac{\kappa}{2}
+{3\over4}\overline{f_{\rm T}}\Big)\delta p^{[\phi]}\nonumber\\
\approx &\overline{f_{\rm Q}}\(4H\dot{\Psi}+3H^2\Psi+\dot{H}\Psi+\ddot{\Psi}\)\nonumber\\
& +12H\overline{f_{\rm QQ}}\(4H^2\dot{\Psi}+3H^3\Psi+4H\dot{H}\Psi+3\dot{H}\dot{\Psi}+\ddot{\Psi}\)\nonumber\\
&+12H^2\dot{\overline{f_{\rm QQ}}}\(H\Psi+\dot{\Psi}\),
\end{align}
\begin{align}\label{46}
&\(\overline{\rho^{[\phi]}}+\overline{p^{[\phi]}}\)\(\kappa+\overline{f_{\rm T}}\)a(t)\delta {\mathfrak u}\nonumber\\
&\approx 2\overline{f_{\rm
Q}}\(H\Psi+\dot{\Psi}\)+12H\overline{f_{\rm
QQ}}\(2\dot{H}\Psi+H\dot{\Psi}\),
\end{align}
where $\overline{f_{\rm QQ}}=\partial \overline{f_{\rm
Q}}/\partial \overline{Q}$ and $\overline{f_{\rm TT}}=\partial
\overline{f_{\rm T}}/\partial \overline{T}$. Note that, Eqs.
\eqref{44}-\eqref{46} are reduced to the corresponding equations
in GR~\cite{Weinberg} when $\overline{f_{\rm Q}}=-1$,
$\overline{f_{\rm QQ}}=0$ and $\overline{f_{\rm
T}}=0=\overline{f_{\rm TT}}$.

In addition, by substituting Eq. \eqref{covleviT} into Eq.
\eqref{75}, for the case of warm inflation, one obviously obtains
\begin{align}\label{covleviTwarm}
&-\(\overline{f_{\rm T}}+\kappa\)\overline{{\cal D}}_{\mu}
\overline{T^{[\phi]}}^{\mu}{}_{\nu}+ \overline{f_{\rm T}}
\partial_\nu \overline{p^{[\phi]}}
-\dfrac{1}{2}\overline{f_{\rm T}} \partial_\nu \overline{T^{[\phi]}}\nonumber\\
&=\overline{B}_\nu+\(\overline{f_{\rm T}}+\kappa\)
\overline{\Lambda}_\nu\equiv \overline{U}_\nu.
\end{align}
The perturbation of the right-hand side of this equation gives
\begin{align}\label{deltaU}
&\delta U_\mu=\delta B_\mu+\delta f_{\rm T} \overline{\Lambda}_\mu+\(\overline{f_{\rm T}}+\kappa\) \delta \Lambda_\mu\nonumber\\
&\hspace*{0.62cm}=\delta B_\mu+\overline{ f_{\rm TT}}\,
\overline{\Lambda}_\mu \delta T^{[\phi]}
+\(\overline{f_{\rm T}}+\kappa\) \delta \Lambda_\mu\nonumber\\
&\hspace*{0.62cm}=\delta B_\mu+ \overline{f_{\rm TT}}\,
\overline{\Lambda}_\mu\(\!-\delta \rho^{[\phi]}+3\delta
p^{[\phi]}\)\! +\(\overline{f_{\rm T}}+\kappa\) \delta
\Lambda_\mu,
\end{align}
where we have assumed that the function $f(Q,T)$ does~not have
mixed terms in $Q$ and $T$, i.e. $f(Q,T)=g(Q)+h(T)$. Then, the
0-component of the perturbation of Eq. \eqref{covleviTwarm}
reads~\cite{Antonio2022}
\begin{align}\label{47}
&\Bigg[\kappa+\overline{f_{\rm T}}+{1\over2}\overline{f_{\rm
T}}\(1-{\partial_0\!\(\delta p^{[\phi]}\)
\over\partial_0\!\(\delta\rho^{[\phi]}\)}\)\Bigg]\dot{\delta} \nonumber\\
&\!-\!\Bigg\lbrace\! 3H\!\(\kappa+\overline{f_{\rm
T}}\)\!\Bigg[\!{\(\kappa\!+\!\overline{f_{\rm
T}}\)w^{[\phi]}\!-\!{1\over2}\overline{f_{\rm
T}}\(1\!-\!w^{[\phi]}\) \over \kappa+\overline{f_{\rm
T}}+{1\over2}\overline{f_{\rm T}}\(1-w^{[\phi]}\)}
\!-\!{\delta p^{[\phi]}\over\delta\rho^{[\phi]}}\!\Bigg]\nonumber\\
&+{3H\(\kappa+\overline{f_{\rm T}}\)(w^{[\phi]}+1)\over
2\Big[\kappa+\overline{f_{\rm T}}+{1\over2}\overline{f_{\rm
T}}\(1-w^{[\phi]}\)\Big]} \Bigg[\overline{f_{\rm
T}}\(1-{\partial_0\!\(\delta p^{[\phi]}\)
\over\partial_0\!\(\delta\rho^{[\phi]}\)}\) \nonumber\\
&+\overline{f_{\rm TT}}\overline{\rho^{[\phi]}}\(1-3{\delta
p^{[\phi]}\over\delta\rho^{[\phi]}}\)\(1-3w^{[\phi]}\)\Bigg]
+3H\overline{f_{\rm TT}}\overline{\rho^{[\phi]}}\nonumber\\
&\times \Big(1-3{\delta
p^{[\phi]}\over\delta\rho^{[\phi]}}\Big)(w^{[\phi]}+1)\Bigg[{\overline{f_{\rm
T}}\(1-w^{[\phi]}\)\over 2\kappa
+\overline{f_{\rm T}}\(3-w^{[\phi]}\)}\Bigg]\Bigg\rbrace \delta \nonumber\\
&+ \(\kappa+\overline{f_{\rm
T}}\)\frac{\overline{\rho^{[\phi]}}+\overline{p^{[\phi]}}}{\overline{\rho^{[\phi]}}}
\bigg[a^{-1}\nabla^2(\delta {\mathfrak u})-3\dot{\Psi}\bigg]\nonumber\\
&\approx \dfrac{-\nabla^2\left[8H\overline{f_{\rm
Q}}\Psi-12H^2\overline{f_{\rm QQ}}\(\dot{\Psi}
+H\Psi\)\right]}{a^2\overline{\rho^{[\phi]}}}\nonumber\\
&\hspace*{0.5cm}+\dfrac{\overline{f_{\rm TT}}\(-\delta
\rho^{[\phi]}+3\delta
p^{[\phi]}\)\overline{\Lambda}_0+\(\overline{f_{\rm
T}}+\kappa\)\delta \Lambda_0}{\overline{\rho^{[\phi]}}},
\end{align}
where
$\left(\rho^{[\phi]}-\overline{\rho^{[\phi]}}\right)/\overline{\rho^{[\phi]}}
=\delta\rho^{[\phi]}/\overline{\rho^{[\phi]}}\equiv\delta$ is the
density contrast parameter,
$w^{[\phi]}={\overline{p^{[\phi]}}/\overline{\rho^{[\phi]}}}$
represents the equation of state parameter, and as $w^{[\phi]}$ is
a constant, we have used\footnote{According to the Newton-Laplace
equation, the speed of sound in fluids, in general, is
$c_s^2=\(\partial p/\partial \rho\)_S $, where the last $S$ stands
for isentropically, i.e. constant entropy. Hence, in the
literature, $c_s^2=\dot{p}/\dot{\rho}$ has been used in, e.g.,
Refs.~\cite{Antonio2022,Pati2022, Bhagat2023}. }
$\dot{\overline{p^{[\phi]}}}/\dot{\overline{\rho^{[\phi]}}}=w^{[\phi]}$.
We recall that the right-hand side of this equation comes from
$\delta U_\mu$ which is due to nonmetricity and warm inflation.
Hence, in the case of metricity and without warm inflation, in the
slow-roll regime by taking $w^{[\phi]}\approx -1$, Eq. \eqref{47}
gives
\begin{align}\label{New47}
&\Bigg[\kappa+\overline{f_{\rm T}}+{1\over2}\overline{f_{\rm
T}}\(1-{\partial_0\(\delta p^{[\phi]}\)
\over\partial_0\(\delta\rho^{[\phi]}\)}\)\Bigg]\dot{\delta} \nonumber\\
&+3H\(\kappa+\overline{f_{\rm
T}}\)\frac{\delta\rho^{[\phi]}+\delta
p^{[\phi]}}{\overline{\rho^{[\phi]}}}
\nonumber\\
&+ \(\kappa+\overline{f_{\rm
T}}\)\frac{\overline{\rho^{[\phi]}}+\overline{p^{[\phi]}}}{\overline{\rho^{[\phi]}}}
\bigg[a^{-1}\nabla^2(\delta {\mathfrak
u})-3\dot{\Psi}\bigg]\approx 0.
\end{align}
This equation recalls the corresponding equation in
GR~\cite{Weinberg} when $\overline{f_{\rm T}}=0$.

On the other hand, by using relations \eqref{rho-p},
\eqref{upMetricPerturb} and $\delta V=\overline{V}'\delta\phi$,
while paying attention to relation \eqref{inflatonlagrangy} to
know where $(\dot{\overline{\phi}})^2$ comes from, the
perturbation of the energy density and pressure density of the
scalar field is obtained as
\begin{align}
& \label{48.1} \delta\rho^{[\phi]}\approx \dot{\overline{\phi}}\,\partial_0(\delta\phi)
+ \overline{V}'\(\overline{\phi}\)\delta\phi - (\dot{\overline{\phi}})^2\Psi, \\
& \label{48.2} \delta p^{[\phi]} \approx
\dot{\overline{\phi}}\,\partial_0(\delta\phi) -
\overline{V}'\(\overline{\phi}\)\delta\phi -
(\dot{\overline{\phi}})^2\Psi.
\end{align}
The sum of these equations clearly gives
\begin{equation}\label{deltaRho-p}
\delta\rho^{[\phi]}+\delta p^{[\phi]}\approx
2\[\dot{\overline{\phi}}\,\partial_0(\delta\phi) -
(\dot{\overline{\phi}})^2\Psi\].
\end{equation}

Up to here, the main unknowns of perturbation are
$\Psi=\Psi(t,{\bf x}) $, $\delta {\mathfrak u}=\delta {\mathfrak
u}(t,{\bf x}) $, $\delta \phi=\delta \phi(t,{\bf x})$, $\delta
\rho^{[\gamma]}=\delta \rho^{[\gamma]}(t,{\bf x})$, $\delta
\rho^{[\phi]}=\delta \rho^{[\phi]}(t,{\bf x})$ and $\delta
p^{[\phi]}=\delta p^{[\phi]}(t,{\bf x})$, which are supposed to be
determined through the model equations. Meanwhile, as mentioned,
the other quantities involved will be perturbed subsequently, e.g.
$V\equiv \overline{V}(\phi)=\overline{V}\(\overline{\phi}+\delta
\phi\)\approx
\overline{V}\(\overline{\phi}\)+\overline{V}'\(\overline{\phi}\)\delta
\phi$, and in the same way,
$\rho^{[\phi]}\approx\overline{\rho^{[\phi]}}+\overline{\rho^{[\phi]}}'\delta
\phi$ and $\Upsilon \approx\overline{\Upsilon}
+\overline{\Upsilon}'\delta\phi $ when $\Upsilon$ is an explicit
function of $\phi$. However, later we will consider special forms
for the potential and the dissipation coefficient. Accordingly, we
have a system of six coupled differential equations
(\ref{38})-(\ref{46}) and \eqref{47} for the unknown quantities.
In general, these equations enable us to explore the evolution of
perturbed quantities. However, up to here, mentioning the
approximate equations is due to their writing up to the
second-order of perturbations. Since each quantity in the
background and perturbation varies slowly during warm inflation,
in the continuation, we rewrite all the relevant equations
according to the corresponding slow-roll approximations.

To proceed, we first transfer all the perturbed quantities into
Fourier space, which results in the spatial components of the
quantities taking the form $e^{i{\bf k}.{\bf x}}$, where ${\bf k}$
represents the wavenumber vector of the corresponding modes.
Accordingly, in this process, we introduce the following
definitions\footnote{In the literature, $\delta {\mathfrak u}(t)$
has been introduced as proportional to $v(t)/k$; however, we
prefer to keep it consistent with the other relevant quantities.}
\begin{eqnarray}\label{deltaPsi-u}
&& \Psi (t,{\bf x})=\Psi(t) e^{i{\bf k}.{\bf x}}, \qquad\qquad\
\delta {\mathfrak u}(t,{\bf x})=\delta {\mathfrak u}(t) e^{i{\bf
k}.{\bf x}},\cr
 &&\delta \phi(t,{\bf x})=\delta\phi(t)e^{i{\bf k}.{\bf
 x}},\qquad\ \quad\
 \delta \rho^{[\gamma]}(t,{\bf x})=\delta \rho^{[\gamma]}(t) e^{i{\bf k}.{\bf
 x}},\cr
 &&\!\delta \rho^{[\phi]}(t,{\bf x})=\delta\rho^{[\phi]}(t)e^{i{\bf k}.{\bf x}},
 \qquad
 \delta p^{[\phi]}(t,{\bf x})=\delta p^{[\phi]}(t) e^{i{\bf k}.{\bf
 x}}.\cr &
\end{eqnarray}
Therefore, we can substitute $\partial_j\rightarrow ik_j$ and
$\nabla^2\rightarrow -k^2$ wherever necessary in the system
equations.

Then, for the purpose of slow-roll approximations, we eliminate
any quantity with a higher time derivative and assume the
plausible approximations $\dot{H}\ll H^2$, $\ddot{\Psi}\ll
H\dot{\Psi}$, $\dot{\Psi}\ll H\Psi$, $\dot{H}\dot{\Psi}\ll
H^3\Psi$ and $\dot{\overline{f_{\rm QQ}}}\ll H\overline{f_{\rm
QQ}}$. Also, we consider modes whose longest wavelength satisfies
${k/a(t)}\ll H$. Hence, under these considerations, Eqs.
\eqref{44} and \eqref{45} are reduced, respectively, to
\begin{align}\label{00deleqar}
&\delta \rho^{[\phi]}(t)\left[\kappa+\dfrac{3}{2}\overline{f_{\rm
T}}-\overline{f_{\rm TT}}\left(\overline{\rho^{[\phi]}}
+\overline{p^{[\phi]}}\right)\right]\nonumber\\
&+\delta p^{[\phi]}(t)\left[-\dfrac{1}{2}\overline{f_{\rm
T}}+3\overline{f_{\rm TT}}\left(\overline{\rho^{[\phi]}}
+\overline{p^{[\phi]}}\right)\right]\nonumber\\
&\approx 6H^2\Psi(t)\left(\overline{f_{\rm Q}}+12 \overline{f_{\rm
QQ}}H^2\right),
\end{align}
\begin{align}\label{tracedeleq}
 &\dfrac{1}{2}\overline{f_{\rm T}}\delta
\rho^{[\phi]}(t)-\left(\!\kappa+\dfrac{3}{2}\overline{f_{\rm
T}}\right)\delta p^{[\phi]}(t)\nonumber\\
 &\approx 6H^2 \Psi(t)
\left(\overline{f_{\rm Q}}+12 \overline{f_{\rm QQ}}H^2\right).
\end{align}
We also assume\footnote{Of course, this is a restrictive
assumption.}\
 that the order of $\overline{f_{\rm T}}$ is the
same as the order of $\overline{T^{[\phi]}}\,\overline{f_{\rm
TT}}$. Then, using relation~\eqref{TraceT} and the slow-roll
approximation $(\dot{\overline{\phi}})^2 \ll \overline{V}$, we get
$\overline{f_{\rm
TT}}\(\overline{\rho^{[\phi]}}+\overline{p^{[\phi]}}\) \ll
\overline{f_{\rm T}}$. Hence, Eq. \eqref{00deleqar} reads
\begin{align}\label{NEWtracedeleq}
 &\left(\kappa+\dfrac{3}{2}\overline{f_{\rm T}}\right)\delta
\rho^{[\phi]}(t)-\dfrac{1}{2}\overline{f_{\rm T}} \delta
p^{[\phi]}(t)\nonumber\\
 &\approx 6H^2 \Psi(t) \left(\overline{f_{\rm
Q}}+12 \overline{f_{\rm QQ}}H^2\right).
\end{align}
Thereafter, subtracting Eq. \eqref{tracedeleq} from Eq.
\eqref{NEWtracedeleq} gives
\begin{equation}\label{rhopzero}
 \delta \rho^{[\phi]}(t)+\delta p^{[\phi]}(t)\approx 0,
\end{equation}
and in turn, $w^{[\phi]}\approx -1$, as expected during inflation.
Accordingly, first inserting relations (\ref{deltaPsi-u}) into Eq.
\eqref{deltaRho-p} and then substituting its result into Eq.
\eqref{rhopzero} yields
\begin{equation}\label{Psi1}
\Psi(t)\approx\dfrac{\partial_0 [\delta \phi(t)]
}{\dot{\overline{\phi}}},
\end{equation}
and in turn, Eqs. \eqref{48.1} and \eqref{48.2} give
\begin{equation}\label{RhoAnd-p}
 \delta \rho^{[\phi]}(t)\approx \overline{V}'\(\overline{\phi}\)\delta\phi(t)\approx -\delta p^{[\phi]}(t).
\end{equation}

Similarly, we assume that the order of $\overline{f_{\rm Q}}$ is
the same as the order of $Q\,\overline{f_{\rm QQ}}$. Then, using
relation~\eqref{QasH} and the slow-roll approximation $\dot{H}\ll
H^2$, we get $\overline{f_{\rm QQ}} \dot{H}\ll \overline{f_{\rm
Q}}$. Hence, while also using the slow-roll condition
$\dot{\Psi}\ll H\Psi$, Eq. \eqref{46} is reduced to
\begin{equation}
\label{deltaueq1}\delta {\mathfrak
u}(t)\approx\dfrac{2\overline{f_{\rm Q}}
H\Psi(t)}{a(t)\(\overline{\rho^{[\phi]}}+\overline{p^{[\phi]}}\)\(\kappa+\overline{f_{\rm
T}}\)}.
\end{equation}
Furthermore, by using the slow-roll approximations $\partial_0
(\delta \rho^{[\gamma]})\ll H \delta \rho^{[\gamma]}$,
$\dot{\Psi}\ll H\Psi$ and $k/a(t)\ll H$,\footnote{Actually, for
adequately slow evolution, we ignore terms involving
long-wavelength perturbations.}\
 and employing Eq.
\eqref{bbradiation} and relation $\delta
\Upsilon=\overline{\Upsilon}^{\prime} \delta\phi$, Eq. \eqref{38}
is reduced to
\begin{equation}\label{51}
\dfrac{\delta
\rho^{[\gamma]}(t)}{\overline{\rho^{[\gamma]}}}\!\approx\!
\dfrac{\overline{\Upsilon}^\prime}{\overline{\Upsilon}}\delta\phi(t)
-\Psi(t)+2\dfrac{\partial_0\[\delta\phi(t)\]}{\dot{\overline{\phi}}}\!\approx\!\dfrac{\overline{\Upsilon}^\prime}
{\overline{\Upsilon}}\delta\phi(t)+\Psi(t),
\end{equation}
where we have used relation \eqref{Psi1} in the last
approximation. Also, by transferring to momentum space, Eq.
(\ref{38.2}) reads
\begin{align}\label{43}
&\overline{\rho^{[\gamma]}}\partial_0\[\delta {\mathfrak
u}(t)\]+\dot{\overline{\rho^{[\gamma]}}}\delta {\mathfrak
u}(t)+4H\overline{\rho^{[\gamma]}} \delta {\mathfrak
u}(t)\nonumber\\
&\approx -\frac{1}{a(t)}
\left[{3\over4}\overline{\Upsilon}\,\dot{\overline{\phi}}\delta\phi(t)
+{\delta\rho^{[\gamma]}(t)\over
4}+\overline{\rho^{[\gamma]}}\Psi(t)\right].
\end{align}
Then, by applying the slow-roll approximation $\partial_0\left[
\overline{\rho^{[\gamma]}}\delta {\mathfrak u}(t)\right]\ll H
\overline{\rho^{[\gamma]}}\delta {\mathfrak u}(t)$ while using Eq.
\eqref{51}, it reads
\begin{equation}\label{52.1}
\delta {\mathfrak u}(t)\approx
-\dfrac{1}{4aH}\left[\dfrac{3\overline{\Upsilon}\,\dot{\overline{\phi}}}{4\overline{\rho^{[\gamma]}}}\delta\phi(t)
+\dfrac{\overline{\Upsilon}^{\prime}}{4\overline{\Upsilon}}\delta\phi(t)
+ \dfrac{5}{4}\Psi(t)\right].
\end{equation}
Next, by substituting $\delta {\mathfrak u}(t)$ from Eq.
\eqref{deltaueq1} into Eq. \eqref{52.1}, while using relations
\eqref{rho-p} and Eq. \eqref{bbradiation}, we obtain
\begin{equation}\label{54}
\Psi(t)\approx -
\dfrac{\dot{\overline{\phi}}\(\kappa+\overline{f_{\rm
T}}\)\(\overline{\Upsilon}^{\prime}\,\dot{\overline{\phi}}
+12H\overline{\Upsilon}\)}{\overline{\Upsilon}\Big[5(\dot{\overline{\phi}})^2\(\kappa+\overline{f_{\rm
T}}\)+32H^2\overline{f_{\rm Q}}\Big]} \delta\phi(t).
\end{equation}

Finally, while considering Eq. \eqref{New47}, and using Eqs.
\eqref{rho-pApproxi} and \eqref{rhopzero}, and the slow-roll
condition $\dot{\Psi}\ll H\Psi$, Eq. \eqref{47} is reduced to
\begin{align}\label{55}
&\frac{\(\kappa+2\overline{f_{\rm
T}}\)}{\overline{\rho^{[\phi]}}(t)}
\[\partial_0\!\(\delta\rho^{[\phi]}(t)\)-\frac{\dot{\overline{\rho^{[\phi]}}}(t)}
{\overline{\rho^{[\phi]}}(t)}\delta\rho^{[\phi]}(t)\]\nonumber\\
&\approx
\dfrac{4k^2H\Psi(t)}{a^2(t)\overline{\rho^{[\phi]}}(t)}\(2\overline{f_{\rm
Q}}-3H^2\overline{f_{\rm QQ}}\)-\dfrac{4\overline{f_{\rm
TT}}\delta \rho^{[\phi]}(t)}{\overline{\rho^{[\phi]}}(t)}
\overline{\Lambda}_0\nonumber\\
&~~~+\dfrac{\overline{f_{\rm
T}}+\kappa}{\overline{\rho^{[\phi]}}(t)}\delta \Lambda_0.
\end{align}
At this stage, all the perturbed equations involve exclusively
temporal aspects.

In the continuation, we represent the system equations for the
warm inflation model in the linear version of $f(Q,T)$ gravity,
relation \eqref{linearf}. At first, by applying Eqs. \eqref{h2ap}
and \eqref{appk} into relation \eqref{54}, it reads
\begin{align}
 \Psi(t)\approx &- \dfrac{\overline{V}^\prime \(\kappa+\beta\)\left[\alpha  \overline{V}^\prime \overline{\Upsilon}^\prime
 + 12\overline{\Upsilon}\, \overline{V} \(\kappa+\beta-\kappa\alpha\overline{\chi}\)\right]}
 {\overline{\Upsilon}\left[5\alpha \overline{V}^{\prime 2}
 \(\kappa+\beta\)+32\overline{V}^2\(\kappa+\beta-\kappa\alpha\overline{\chi}\)^2\right]} \nonumber\\
&\label{lastpsi}\times \delta\phi(t).
\end{align}

Thereafter, by substituting relations \eqref{rho-pApproxi},
\eqref{37} (while using relations \eqref{Psi1} and $\delta
\Upsilon=\overline{\Upsilon}^{\prime} \delta\phi$ into it) and
\eqref{RhoAnd-p} into Eq. \eqref{55}, and then rearranging its
terms, we obtain
\begin{align}\label{55New}
&\left(\kappa+2\beta\right)\left[\dfrac{\overline{V}''}{\overline{V}}-\dfrac{\overline{V}'^2}
{\overline{V}^2}+\dfrac{\left(\kappa+\beta\right)\overline{\Upsilon}'\dot{\overline{\phi}}}{\overline{V}\left(\kappa+2\beta\right)}
\right]\delta \phi(t)\nonumber\\
&\approx\!
\left[\!-\dfrac{\left(\kappa+2\beta\right)\overline{V}'}{\overline{V}\,\dot{\overline{\phi}}}\!
-\!\left(\kappa+\beta\right)\dfrac{\overline{\Upsilon}}{\overline{V}}\right]\!\partial_0
\left[ \delta \phi(t)\right]\!+\!\dfrac{8\alpha k^2H\Psi(t)}{a^2
\overline{V}\,\dot{\overline{\phi}}}.
\end{align}
Now, by using first Eq. \eqref{appk} and then Eq. \eqref{h2ap}
into the last term on the right-hand side of this equation, it
reads
\begin{equation}
\dfrac{8\,\alpha k^2H\Psi(t)}{a^2
\overline{V}\,\dot{\overline{\phi}}}\approx \dfrac{8
k^2\left(\kappa+\beta-\kappa\alpha\overline{\chi}\right)}{a^2
\overline{V}'}\Psi(t).
\end{equation}
Thereupon, by substituting $\Psi(t)$ from relation \eqref{lastpsi}
into it, while using the slow-roll parameters \eqref{epsv} and
\eqref{lamv}, we obtain
\begin{align}
&\dfrac{8\alpha k^2H\Psi(t)}{a^2
\overline{V}\,\dot{\overline{\phi}}}\nonumber\\
&\approx\dfrac{-8
k^2\(\kappa+\beta\)\left(\kappa+\beta-\kappa\alpha\overline{\chi}\right)\left(12-\lambda\right)}
{a^2\overline{V}\left[-10\(\kappa+\beta\)\epsilon
+32\left(\kappa+\beta
-\kappa\alpha\overline{\chi}\right)\right]}\delta\phi(t)\nonumber\\
&\label{rhsterm}\approx \dfrac{-3k^2
\(\kappa+\beta\)}{a^2\overline{V}}\delta\phi(t)\approx\frac{k^2}{a^2H^2}\dfrac{
\left(\kappa+\beta\right)\left(\kappa+2\beta\right)}{\alpha
}\delta\phi(t),
\end{align}
where we have employed conditions $\epsilon \ll 1$ and $\lambda\ll
1$ in the middle approximation and Eq. \eqref{h2ap} in the last
approximation. However, due to long-wavelength perturbations, i.e.
with $k^2\ll a^2 H^2$, we can ignore this term. Therefore, by
using definition \eqref{dissipation rate} and Eq. \eqref{appk}
into Eq. \eqref{55New}, it reads
\begin{align}\label{DeltaPhiEq}
&\left(\kappa+2\beta\right)\left[
\dfrac{\overline{V}''}{\overline{V}}-\dfrac{\overline{V}'^2}{\overline{V}^2}-\dfrac{\left(\kappa
+\beta\right)\overline{\chi}}{\left(\kappa+\beta-\kappa\alpha\overline{\chi}\right)}\dfrac{\overline{\Upsilon}'
\overline{V}'}{\overline{\Upsilon}\,\overline{V}}\right]
\delta \phi(t)\nonumber\\
&\approx
\dfrac{3H}{\overline{V}}\Big[\left(\kappa+\beta-\kappa\alpha\overline{\chi}\right)
-\left(\kappa+\beta\right)\overline{\chi}\Big]\partial_0
\left[ \delta \phi(t)\right]\nonumber\\
&\approx\frac{-\overline{V}'\left(\kappa+2\beta\right)\Big[\left(\kappa+\beta-\kappa\alpha\overline{\chi}\right)
-\left(\kappa+\beta\right)\overline{\chi}\Big]}{\overline{V}\left(\kappa+\beta-\kappa\alpha\overline{\chi}\right)}\frac{d\left[
\delta \phi(t)\right]}{d\phi},
\end{align}
where $\kappa+2\beta\neq 0$ has been assumed, and Eq. \eqref{appk}
has been used in the last approximation. At last, from Eq.
\eqref{DeltaPhiEq}, we obtain the solution
\begin{equation}\label{57}
\delta \phi(t)\approx C\exp \left[\zeta(\phi)\right],
\end{equation}
where $C$ is a constant and $\zeta(\phi)$ is
\begin{align}\label{zetaeq}
\zeta(\phi)\equiv & \int \left\lbrace
-\dfrac{\kappa+\beta-\kappa\alpha\overline{\chi}}{\left(\kappa
+\beta-\kappa\alpha\overline{\chi}\right)-\left(\kappa+\beta\right)\overline{\chi}}
\left(\dfrac{\overline{V}''}{\overline{V}'}-\dfrac{\overline{V}'}{\overline{V}}\right)\right.\nonumber\\
&\left.
{}\,\,\,\,\,\,\,\,+\dfrac{\left(\kappa+\beta\right)\overline{\chi}}
{\left(\kappa+\beta-\kappa\alpha\overline{\chi}\right)-\left(\kappa+\beta\right)\overline{\chi}}
\dfrac{\overline{\Upsilon}'}{\overline{\Upsilon}}\right\rbrace
{\rm d}\phi.
\end{align}

It has been shown~\cite{Berera1995,Hall2004,Herrera} that the
density fluctuation derived from solution \eqref{57} transforms as
\begin{equation}\label{59}
\Delta_{\rm R}\approx C\approx\delta \phi(t)\exp[-\zeta(\phi)].
\end{equation}
Actually, in this relation, $\delta\phi(t)$ represents the
fluctuation of the scalar field during the warm inflationary era,
generated by the thermal interaction with the radiation field
as~\cite{Berera2000,Herrera}
\begin{equation}
\left[\delta \phi (t)\right]^2=\dfrac{k_{\rm F}\,\tau}{2 \pi^2},
\end{equation}
where $k_{\rm F}$ is the freeze-out scale. The freeze-out time,
$t_{\rm F}$, is determined when the damping rate (i.e., the
coefficient of the term $\partial_0 \left[ \delta \phi(t)\right]$
in the middle line of relation \eqref{DeltaPhiEq}) falls below the
expansion rate $H$~\cite{Hall2004,Taylor2000}. Hence, the
freeze-out scale is
\begin{equation}\label{kfeq}
k_{\rm F}\approx
\sqrt{\dfrac{3H^2}{\overline{V}}\left[\left(\kappa+\beta-\kappa\alpha\overline{\chi}\right)
-\left(\kappa+\beta\right)\overline{\chi}\right]}.
\end{equation}
Accordingly, relation \eqref{59} undergoes the modification
\begin{align}\label{59New}
\Delta_{\rm R}^2& \approx \dfrac{k_{\rm F}\,\tau}{2 \pi^2}\exp[-2\zeta(\phi)]\nonumber\\
&\approx\dfrac{\tau \exp[-2\zeta(\phi)]}{2 \pi^2}
\sqrt{\dfrac{3H^2}{\overline{V}}
\left[\left(\kappa+\beta-\kappa\alpha\overline{\chi}\right)-\left(\kappa+\beta\right)\overline{\chi}\right]}.
\end{align}
The above relation, derived from the perturbation theory at the
horizon crossing, can be compared with the amplitude of the scalar
power spectrum observed through the CMB radiation data at a pivot
scale. In this regard, the Planck $2018$ data and the WMAP three
years data give the value of the amplitude of the scalar power
spectrum as $\Delta_{\rm R}^2(k_0)\approx {\mathcal O}(10^{-9})$
at the pivot scale $k_0=0.002\ {\rm Mpc^{-1}}$~\cite{WMAP2006,
Planck:2018jri, Montefalcone:2022jfw}.

In the continuation, the scalar spectral index, $n_{\rm S}$,
is~\cite{Encyclo2014}
\begin{equation}\label{SSIndex}
n_{\rm S}-1=\dfrac{{\rm d}\ln \Delta_{\rm R}^2}{{\rm d}\ln k},
\end{equation}
where the inflationary observables are calculated at the first
horizon crossing~\cite{Encyclo2014}, i.e. when $k=aH$. Hence, by
inserting relation \eqref{59New} into relation \eqref{SSIndex}, we
get
\begin{align}\label{nsk}
n_{\rm S}-1& \approx\dfrac{1}{2}\dfrac{{\rm d}\ln k_{\rm
F}^2}{{\rm d}\ln (aH)}+\dfrac{1}{4}\dfrac{{\rm d}\ln \tau^4}{{\rm
d}\ln (aH)}-\dfrac{2{\rm d}\zeta(\phi)}{{\rm d}\ln (aH)}\nonumber\\
&=\(\dfrac{1}{2}\dfrac{{\rm d}\ln k_{\rm F}^2}{{\rm
d}\phi}+\dfrac{1}{4}\dfrac{{\rm d}\ln \tau^4}{{\rm
d}\phi}-\dfrac{2{\rm d}\zeta(\phi)}{{\rm
d}\phi}\)\dot{\overline{\phi}}\dfrac{{\rm d}t}{{\rm d}\ln (aH)}.
\end{align}
In this regard, by employing relations \eqref{bbradiation},
\eqref{eps1}, \eqref{h2ap}, \eqref{appk} and \eqref{kfeq}, we
obtain
\begin{equation}\label{dlnkffinal}
\dfrac{{\rm d}\ln k_{\rm F}^2}{{\rm
d}\phi}\approx\dfrac{-\left(\kappa+\beta+\kappa\alpha\right)\overline{\chi}}{\left[\left(\kappa
+\beta-\kappa\alpha\overline{\chi}\right)
-\left(\kappa+\beta\right)\overline{\chi}\right]}
\left(\dfrac{\overline{\Upsilon}'}{\overline{\Upsilon}}-\dfrac{1}{2}\dfrac{\overline{V}'}{\overline{V}}\right),
\end{equation}
\begin{align}\label{dlntau4}
\dfrac{{\rm d}\ln \tau^4}{{\rm d}\phi}\!\approx\!
\dfrac{\left(\kappa\!+\!\beta\!+\!\kappa\alpha\overline{\chi}\right)}{\left(\kappa\!+\!\beta\!
-\!\kappa\alpha\overline{\chi}\right)}
\dfrac{\overline{\Upsilon}'}{\overline{\Upsilon}}\!-\!\dfrac{\left(3\kappa\!+\!3\beta\!-\!\kappa\alpha\overline{\chi}\right)}
{2\left(\kappa\!+\!\beta\!-\!\kappa\alpha\overline{\chi}\right)}\dfrac{\overline{V}'}{\overline{V}}\!
+\!2\dfrac{\overline{V}''}{\overline{V}'},
\end{align}
\begin{equation}\label{dlnk}
\dfrac{1}{\dot{\overline{\phi}}}\dfrac{{\rm d}\ln (aH)}{{\rm d}t}
\approx\dfrac{\left(\kappa+\beta-\kappa\alpha\overline{\chi}\right)\overline{V}}{\alpha
\overline{V}'}\left(1-\epsilon\right).
\end{equation}
Now, utilizing these relations, while employing relation
\eqref{zetaeq} and using the modified slow-roll parameters
\eqref{epsv}, \eqref{etv} and \eqref{lamv}, relation \eqref{nsk}
becomes
\begin{align}\label{nseq}
n_{\rm S}-1\approx &\left\lbrace
\dfrac{8\left(\kappa+\beta\right)-\left(\kappa+\beta+9\kappa\alpha\right)\overline{\chi}}
{2\left[\left(\kappa+\beta-\kappa\alpha\overline{\chi}\right)-\left(\kappa+\beta\right)
\overline{\chi}\right]}\right.\nonumber\\
&\left.
~~+\dfrac{3\kappa+3\beta-\kappa\alpha\overline{\chi}}{4\left(\kappa+\beta-\kappa\alpha\overline{\chi}\right)}
\right\rbrace \epsilon\nonumber\\
&-\left\lbrace
\dfrac{1}{2}+\dfrac{2\left(\kappa+\beta-\kappa\alpha\overline{\chi}\right)}
{\left(\kappa+\beta-\kappa\alpha\overline{\chi}\right)
-\left(\kappa+\beta\right)\overline{\chi}}\right\rbrace \eta\nonumber\\
&+\left\lbrace
\dfrac{\left(5\kappa+5\beta+\kappa\alpha\right)\overline{\chi}}
{2\left[\left(\kappa+\beta-\kappa\alpha\overline{\chi}\right)
-\left(\kappa+\beta\right)\overline{\chi}\right]}\right.\nonumber\\
&\left.
~~~~-\dfrac{\kappa+\beta+\kappa\alpha\overline{\chi}}{4\left(\kappa
+\beta-\kappa\alpha\overline{\chi}\right)}\right\rbrace \lambda,
\end{align}
where condition $1 + \epsilon \approx 1$ has been used.

\subsection{TENSOR PERTURBATIONS}
In this subsection, we consider the evolution of tensor
perturbations. For this purpose, we denote the related metric
as~\cite{Bhattacharyaa2006}
\begin{equation}\label{63}
ds^2=-dt^2+a^2(t)(\delta_{ij}+h_{ij}) dx^i dx^j,
\end{equation}
wherein tensor perturbations are characterized by the spatially
symmetric, transverse, traceless tensor $h_{ij}$, with properties
such that $\partial^ih_{ij}=0=h^i_i$. Accordingly, the result can
be achieved by perturbing the energy-momentum tensor up to the
second-order within the framework of linear perturbations
as~\cite{Antonio2022}
\begin{equation}\label{64}
\ddot{h}_{ij}+\Big\lbrace
3H-\dfrac{d}{dt}\Big[\ln(f_Q)\Big]\Big\rbrace
\dot{h}_{ij}+\dfrac{k^2}{a^2} h_{ij}=0.
\end{equation}
This equation determines the evolution of tensor perturbations. It
is interesting to note that the functional form of Eq.~\eqref{64}
is the same as the corresponding one obtained in $f(Q)$
gravity~\cite{Jimenez2020}. Still, in the linear functional form
of $f(Q,T)$ gravity, relation \eqref{linearf}, one has
$d[\ln(f_Q)]/dt=0$. Nevertheless, the contribution of $f(Q,T)$
gravity from the source $T^{[\phi]}$ to the tensor modes can be
obtained through the presence of the Hubble parameter term, $H$,
in Eq.~\eqref{64}. Indeed, the modified Friedmann equations of the
model, specifically Eqs.~\eqref{rho11} and \eqref{p11}, determine
the effect of the parameters of such a model on the tensor
perturbations.

In Ref.~\cite{Bhattacharya2006}, a method has been provided to
compute the emergence of tensor perturbations during warm
inflation by introducing an additional factor
$\coth\left[k/(2\,\tau)\right]$ that modifies the power spectrum
of the tensor fluctuation modes as~\cite{Herrera}
\begin{equation}\label{65}
\Delta_{\rm T}^2={4H^2\over \pi}\coth\(\dfrac{k}{2\,\tau}\).
\end{equation}
Utilizing Eq. \eqref{h2ap}, this relation becomes
\begin{equation}\label{66}
\Delta_{\rm T}^2\approx {4(\kappa+2\beta)\overline{V}(\phi)\over
-3\alpha\, \pi}\coth\(\dfrac{k}{2\,\tau}\).
\end{equation}
However, due to the weak nature of gravitational interactions, the
tensor primordial power spectrum remains unaltered, similar to its
state during cold inflation without any modification.

Furthermore, the inflationary parameter, denoted as the
tensor-to-scalar ratio $r$, is
\begin{equation}
{\rm r}=\dfrac{\Delta_{\rm T}^2}{\Delta_{\rm R}^2}.
\end{equation}
By applying relations \eqref{59New} (while using Eq. \eqref{h2ap}
into it) and \eqref{66}, it reads
\begin{equation}\label{rstrong}
r\approx \dfrac{\kappa\,\overline{V}(\phi)\sqrt{(\kappa+2\beta)}
\exp\[2\zeta(\phi)\]}{3\,\tau \sqrt{-\alpha
\left[\left(\kappa+\beta-\kappa\alpha\overline{\chi}\right)-\left(\kappa
+\beta\right)\overline{\chi}\right]}} \coth\(\dfrac{k}{2\,\tau}\),
\end{equation}
which dictates
\begin{equation}\label{ConsOnKapaBeta}
\kappa+2\beta >0
\end{equation}
and
\begin{equation}\label{ConsOnKapaBetaKhiAlpha}
\kappa+\beta-\kappa\alpha\overline{\chi}> \left(\kappa
+\beta\right)\overline{\chi}
\end{equation}
for negative values of $\alpha $.

In the warm inflationary paradigm, the primordial scalar power
spectrum undergoes an increase during the warm inflation era
compared to the cold inflation era, whereas the tensor power
spectrum remains consistent with the cold inflation period.
Consequently, the reduction of the tensor-to-scalar ratio,
compared to the cold inflation scenario, has the potential to
revive models that have been disregarded due to discrepancies with
observational data. Subsequently, we investigate the inflationary
characteristics within the framework of warm inflation,
considering the linear version of $f(Q,T)$ gravity, which is
coupled with both the inflaton field and the radiation field. This
analysis encompasses a power-law potential conducted under the
context of the strong dissipation regime. Within this regime, it
becomes possible to employ the approximation where $\chi\gg 1$.

\section{Strong dissipation regime}
Henceforth, since the quantities involved are unperturbed ones, we
will~not label them with a bar for simplicity. In the strong
dissipation regime, we make the plausible assumption
\begin{equation}\label{conseq}
\beta\ll -\kappa\alpha\chi,
\end{equation}
which, for negative values of $\alpha $, while considering
condition \eqref{NEWconstrainteq}, requires
\begin{equation}\label{conseqNEW}
-\alpha\chi > 1.
\end{equation}
However, we prefer to assume $|-\alpha\chi|\gg 1$. Under these
assumptions, we can write
\begin{equation}\label{strongConditions}
\kappa+\beta-\kappa\alpha\chi\approx -\kappa\alpha\chi.
\end{equation}
Hence, condition \eqref{constrainteq} is always satisfied, since
we have
\begin{equation}
\dfrac{-\alpha}{\kappa+\beta-\kappa\alpha\chi}>0
\quad\longrightarrow\quad\dfrac{1}{\kappa\chi}>0,
\end{equation}
where $\kappa\chi$ is obviously a positive value. Also, for
negative values of $\alpha $, under these assumptions, condition
\eqref{ConsOnKapaBetaKhiAlpha} requires
\begin{equation}\label{NEWconsOnKapaBetaAlpha}
\kappa+\beta+\kappa\alpha < 0.
\end{equation}
In addition, assumption \eqref{conseq} implies that when $\alpha$
takes on negative values, the $\beta$ parameter encompasses both
positive and negative values. Nevertheless, when $\alpha$ is a
positive value, a prerequisite emerges for $\beta$ to hold
negative values.

Furthermore, in the context of the strong dissipation regime and
under these assumptions, the modified slow-roll parameters
\eqref{epsv}, \eqref{etv} and \eqref{lamv} become
\begin{align}\label{srstrongall}
&\epsilon\approx\dfrac{1}{2\kappa\chi}\(\dfrac{V^\prime}{V}\)^2,
\qquad \eta\approx\dfrac{1}
{\kappa\chi}\(\dfrac{V^{\prime\prime}}{V}\),\nonumber\\
&\lambda\approx\dfrac{1}{\kappa\chi}\(\dfrac{\Upsilon^\prime
V^\prime}{\Upsilon V}\).
\end{align}
Also, the e-folding number \eqref{nv} reads
\begin{equation}\label{Nstrong}
N\approx\kappa\int_{\phi_{\rm end}}^{\phi}
\chi\dfrac{V}{V^{\prime}}{\rm d}\phi.
\end{equation}
Thus, the model parameters $\alpha$ and $\beta$ do~not appear in
the slow-roll parameters and the e-folding number. Moreover,
relation \eqref{zetaeq} and the scalar spectral index \eqref{nseq}
are now
\begin{align}\label{zetastrong}
\zeta(\phi)\approx & \int \Bigg[
\dfrac{-\kappa\alpha}{\kappa+\beta+\kappa\alpha}
\left(\dfrac{V''}{V'} -\dfrac{V'}{V}\right)\nonumber\\
&~~~~~-\dfrac{\kappa+\beta}{\kappa+\beta+\kappa\alpha}
 \dfrac{\Upsilon'}{\Upsilon}\Bigg] {\rm
 d}\phi,
\end{align}
\begin{align}\label{ssis}
n_{\rm S}-1\approx
&\dfrac{3\kappa+3\beta+19\kappa\alpha}{4\left(\kappa+\beta+\kappa\alpha\right)}\epsilon
-\dfrac{\kappa+\beta+5\kappa\alpha}{2\left(\kappa+\beta+\kappa\alpha\right)}\eta\nonumber\\
&-\dfrac{9\kappa+ 9\beta+
\kappa\alpha}{4\left(\kappa+\beta+\kappa\alpha\right)}\lambda.
\end{align}

As mentioned, the behavior of the dissipation coefficient has a
crucial impact on the evolution of the inflaton field and the
resulting cosmological observables such as the spectrum of the
scalar and tensor perturbations. It is a key parameter that
distinguishes warm inflation from the standard cold inflation
scenario. In the subsequent analysis, we will investigate the
power-law potential under two variations of the dissipation
coefficient, one of them as a constant value and the other as a
function of the inflaton scalar field to the power of two, within
the context of the strong dissipation regime.

\subsection{POWER-LAW POTENTIAL WITH CONSTANT DISSIPATION COEFFICIENT}
In this subsection, with the power-law potential, we assume a
constant dissipation coefficient, say $\Upsilon_0$. Power-law
potentials represent a class of inflaton potentials that exhibit a
straightforward mathematical structure. It contains the potential
energy $V(\phi)$ which is directly proportional to the inflaton
field $\phi$ raised to a fixed power $n$, namely
\begin{equation}\label{powerpotential}
V(\phi)=\nu\phi^n,
\end{equation}
where $n$ is a positive rational number and $\nu$ is a constant of
the model~\cite{Linde1983, Pavluchenko2004}. In this case, the
slow-roll parameters \eqref{srstrongall} become
\begin{equation}
\label{srpc}\epsilon\approx \dfrac{n^2}{2\kappa\chi_0\phi^2},
\qquad \eta\approx\dfrac{n(n-1)}{\kappa\chi_0\phi^2},
\qquad\lambda\approx 0.
\end{equation}
Also, Eq. \eqref{h2ap} reads
\begin{equation}\label{h2power}
3H^2\approx\dfrac{\nu\(\kappa+ 2\beta\)}{-\alpha}\phi^n,
\end{equation}
and in turn, the dissipation rate \eqref{dissipation rate} is now
\begin{equation}\label{chipc}
\chi_0=\dfrac{\Upsilon_0}{3H}\approx \sqrt{\dfrac{-\alpha
\Upsilon_0^2}{3\,\nu\(\kappa+2\beta\)}}\,\phi^{-\dfrac{n}{2}}.
\end{equation}

Furthermore, by considering the condition $\epsilon(\phi_{\rm
end})=1$ alongside the above relations, we obtain the scalar field
at the end of inflation as
\begin{equation}\label{phiendpc}
\phi_{\rm end}\approx\left[\dfrac{3\, \nu
n^4\(\kappa+2\beta\)}{-4\alpha\(\kappa\Upsilon_0\)^2}\right]^{1/(4-n)},
\end{equation}
which dictates $n\neq 4$. Also, the e-folding number
\eqref{Nstrong} reads
\begin{equation}
N\approx \left[\phi^{(4-n)/2}-\phi_{\rm end}^{(4-n)/2}\right]
\left[\dfrac{-4\alpha \(\kappa\Upsilon_0\)^2}{3\nu n^2
\(4-n\)^2\(\kappa+2\beta\)}\right]^{1/2}.
\end{equation}
Subsequently, the scalar field at the first horizon crossing
becomes
\begin{align}\label{phipc}
\phi \approx \left\lbrace \dfrac{-4 \alpha \(\kappa\Upsilon_0\)^2
}{3 \nu n^2 \left(\kappa+2\beta\right) \left[n^2+
\left(4-n\right)^2 N^2\right]}\right\rbrace^{1/\left(n-4\right)}.
\end{align}
In addition, in this case, relation \eqref{zetastrong} is
\begin{equation}\label{zetpc}
\zeta\(\phi\)\approx\int \dfrac{-\kappa\alpha }{\kappa+\beta
+\kappa\alpha }\left(\dfrac{-1}{\phi}\right){\rm d}\phi
=\dfrac{\kappa\alpha }{\kappa+\beta+\kappa\alpha }\ln{\phi}.
\end{equation}

To calculate the temperature of the thermal bath, first by using
relations \eqref{appk}, \eqref{strongConditions} and
\eqref{h2power}, we obtain
\begin{equation}\label{phidotpower}
\dot{\phi}^2\approx\dfrac{ n^2 \nu
\(\kappa+2\beta\)}{-3\,\alpha\kappa^2\chi_0^2}\phi^{(n-2)}.
\end{equation}
Then, from relation~\eqref{bbradiation}, we have
\begin{equation}\label{tauphidot}
\tau=\(\dfrac{3\,\chi_0}{4C_{\gamma}}\dot{\phi}^2\)^{1/4},
\end{equation}
wherein by inserting relations \eqref{chipc} and
\eqref{phidotpower}, it reads
\begin{equation}\label{finaltauconst}
\tau\approx\left[\dfrac{n^2}{4C_{\gamma}\kappa^2\Upsilon_0}\sqrt{\dfrac{3\,\nu^3\(\kappa
+2\beta\)^3}{\(-\alpha\)^3}}\right]^{1/4}\phi^{\dfrac{3n-4}{8}}.
\end{equation}

Thereupon, by substituting relations \eqref{h2ap}, \eqref{chipc},
\eqref{zetpc} and \eqref{finaltauconst} into relation
\eqref{59New} for the strong dissipation regime, we obtain
\begin{align}\label{delta2R}
\Delta_{\rm R}^2 &\approx \sqrt{\dfrac{-n \left(\kappa+\beta+\kappa\alpha\right)}
{4 \pi^4 \kappa}\Big[\dfrac{\nu\Upsilon_0^2\left(\kappa+2\beta\right)^5}
{48 C_{\gamma}^2\left(-\alpha\right)^5}\Big]^{1/4}}\nonumber\\
&~~~\times
\phi^{\dfrac{(n-4)(\kappa+\beta+\kappa\alpha)-16\kappa\alpha}{8(\kappa+\beta+\kappa\alpha)}}.
\end{align}
Then, by inserting relation \eqref{phipc} into the above relation,
we can achieve the density fluctuation in terms of the e-folding
number, $N$, and the other parameters, $n, \nu, \Upsilon_0,
\alpha$ and $\beta$.

Eventually, by substituting relations \eqref{srpc}, while using
relation \eqref{chipc}, into relation \eqref{ssis}, we obtain the
scalar spectral index in terms of the inflaton field as
\begin{align}\label{nspf}
n_{\rm S}-1\approx &\,\sqrt{\dfrac{3\,\nu \left(\kappa+2\beta\right)}{-\alpha\Upsilon_0^2}}\,n\, \phi^{\left(n-4\right)/2}\nonumber\\
&\times \left[\dfrac{4\left(\kappa+\beta+5\kappa\alpha\right)-n
\left(\kappa+\beta+\kappa\alpha\right)}{8\kappa\left(\kappa+\beta+\kappa
\alpha\right)}\right]\nonumber\\
\approx &\dfrac{4\left(\kappa+\beta+5\kappa\alpha\right)-n
\left(\kappa+\beta+\kappa\alpha\right)}{4\left(\kappa+\beta+\kappa
\alpha\right)\sqrt{n^2+ \left(4-n\right)^2 N^2}},
\end{align}
where we have used relation \eqref{phipc} for the last row.
Interestingly, in this model for the case of the strong
dissipation regime, the scalar spectral index is obtained only as
a function of the $\alpha$, $\beta$, $n$ and $N$ parameters and
independent of $\nu$ and $\Upsilon_0$. Also, utilizing relations
\eqref{dissipation rate}, \eqref{strongConditions} and
\eqref{zetpc} into relation \eqref{rstrong}, the tensor-to-scalar
ratio reads
\begin{align}\label{rpc}
r\approx &\dfrac{\kappa\,
\nu}{\tau}\sqrt{\dfrac{H(\kappa+2\beta)}{3\, \alpha
\Upsilon_0\left(\kappa
+\beta+\kappa\alpha\right)}}\coth\(\dfrac{k}{2\tau}\)\nonumber\\
&\times \phi^{\left[\dfrac{n (\kappa+ \beta+\kappa\alpha)
+2\kappa\alpha }{\kappa+\beta+\kappa\alpha}\right]}.
\end{align}
Then, by substituting relations (\ref{phipc}) and
(\ref{finaltauconst}) into relation (\ref{rpc}), we obtain the
inflationary observable $r$ as a function of the $\alpha$,
$\beta$, $n$, $N$, $\Upsilon_0$ and $\nu$ parameters. Of course,
later, when we want to calculate the inflationary observable $r$,
we will use the value of the pivot scale $k_0\approx 0.002\ {\rm
Mpc^{-1}}$ for the wavenumber $k$. Moreover, by substituting
relations (\ref{phipc}) and/or (\ref{phiendpc}) into relation
(\ref{finaltauconst}), we can also obtain the temperature of the
thermal bath at the first horizon crossing and/or at the end of
inflation, respectively.

To evaluate the ability of any theory of gravitation to explain
inflation, it is essential that the predicted values of $n_{\rm
S}$ and $r$ be in agreement with the observational data. These
observables offer valuable insights into the underlying mechanisms
driving inflation and provide a means to test and refine
theoretical models. Achieving a high degree of observational
consistency ensures that our understanding of inflation remains
grounded in empirical evidence and bolsters the credibility of our
scientific endeavors. In this regard, the most recent constraints
established by the Planck collaboration for the scalar spectral
index and the tensor-to-scalar ratio are~\cite{Planck:2018jri}
\begin{align}\label{planckdata}
&n_{\rm S}= 0.9649\pm 0.0042\quad \nonumber\\
&{}\qquad\left({\rm at}\ 68 \%\, {\rm CL, Planck TT,TE,EE+lowE+lensing}\right),\nonumber\\
& r<0.10 \quad\left({\rm at}\ 95 \%\, {\rm CL, Planck
TT+lowE+lensing}\right).
\end{align}
Nevertheless, through the combined examination of the Planck,
BK$15$ and BAO data, additional restrictions have been imposed,
further constricting the upper limit of $r$ to a specific value of
\begin{equation}\label{planckdata2}
r<0.056\quad{\rm at}\ 95 \%\, {\rm CL}.
\end{equation}

On the other hand, in some studies, e.g.,
Ref.~\cite{Montefalcone:2022jfw}, wherein the concept of warm
inflation is examined within the framework of GR for a natural
potential, due to a limited number of parameters in their model,
it is possible to fix the scale of the scalar amplitude to what is
observed in the CMB. Then, based on the number of e-folds, the
parameter of their chosen potential and the dissipation
coefficient can be fixed. However, in this different study, we are
left with many parameters. Moreover, the purpose of this work is
to implement and investigate the impact of the linear version of
$f(Q,T)$ modified gravity on the warm inflationary scenario.
Therefore, to proceed, we are required to keep the parameters of
the potential and the dissipation coefficient fixed with a given
e-folding number. Nevertheless, the scale of scalar amplitudes
provides an additional constraint on the parameters used; hence,
we need to incorporate such a constraint in the obtained results.
In this regard, to ensure the correctness of the values used for
the parameters in drawing the figures, we have also drawn the
density fluctuation in terms of these values. However, our main
goal is to achieve the scalar spectral index and the
tensor-to-scalar ratio obtained from the model in comparison with
the observational data.

\begin{figure*}[t!]
\centering \subfigure[]{
\includegraphics[width=0.31\textwidth]{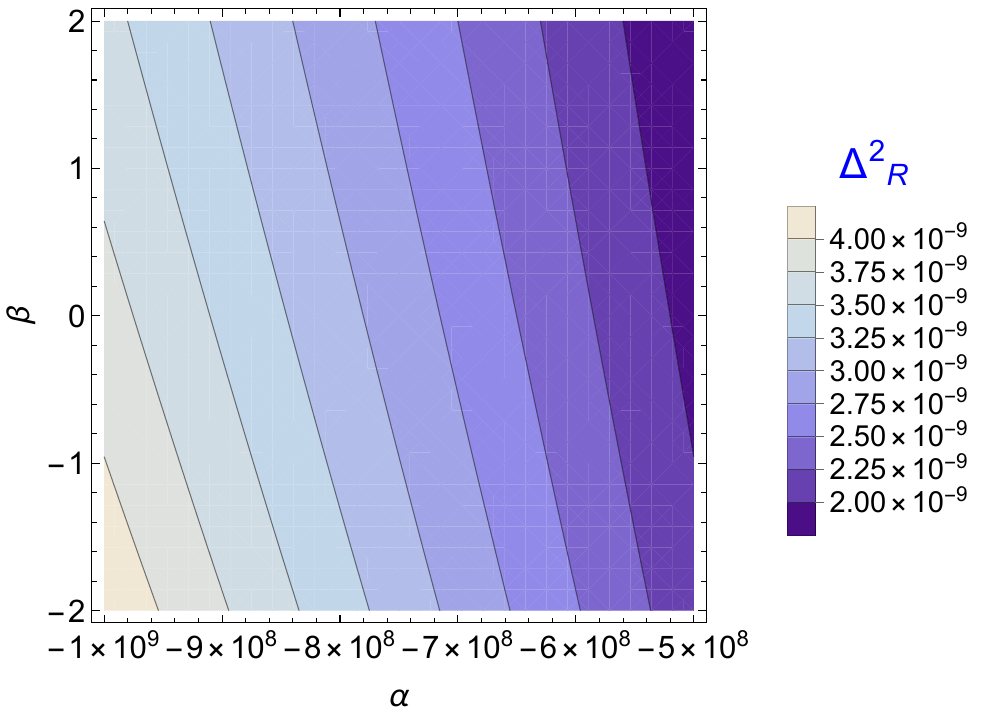}
\label{a1} } \subfigure[] {
\includegraphics[width=0.31\textwidth]{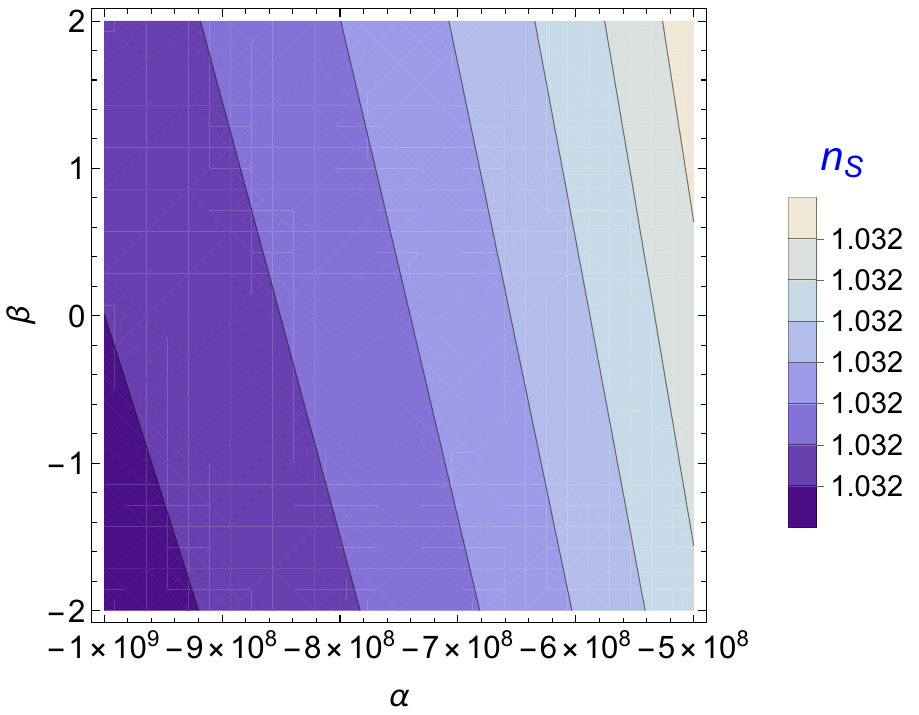}
\label{b1} } \subfigure[] {
\includegraphics[width=0.31\textwidth]{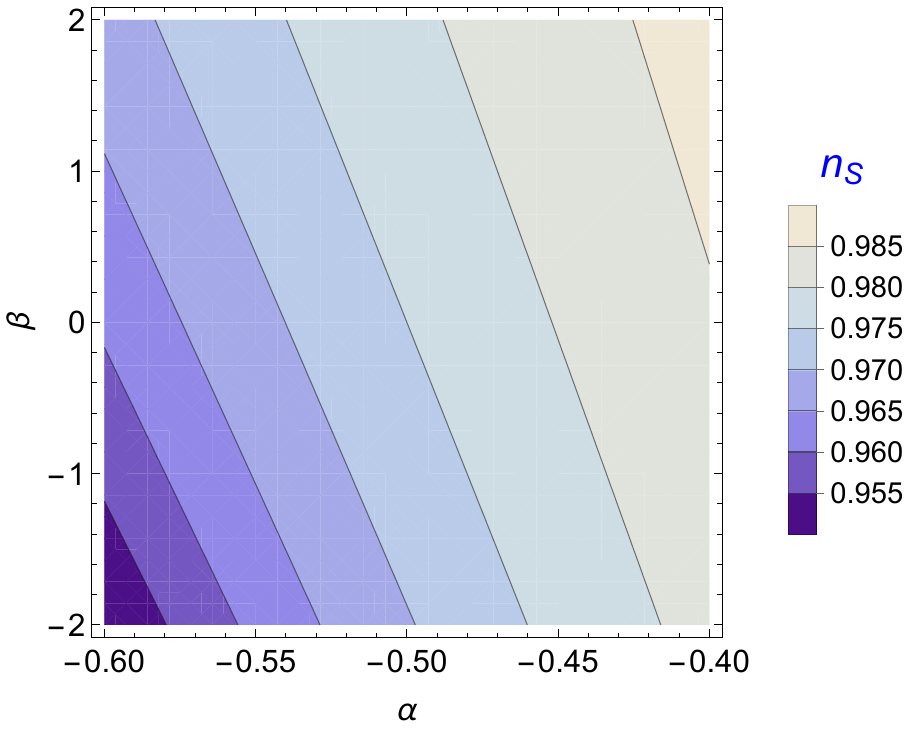}
\label{c1} } \subfigure[] {
\includegraphics[width=0.31\textwidth]{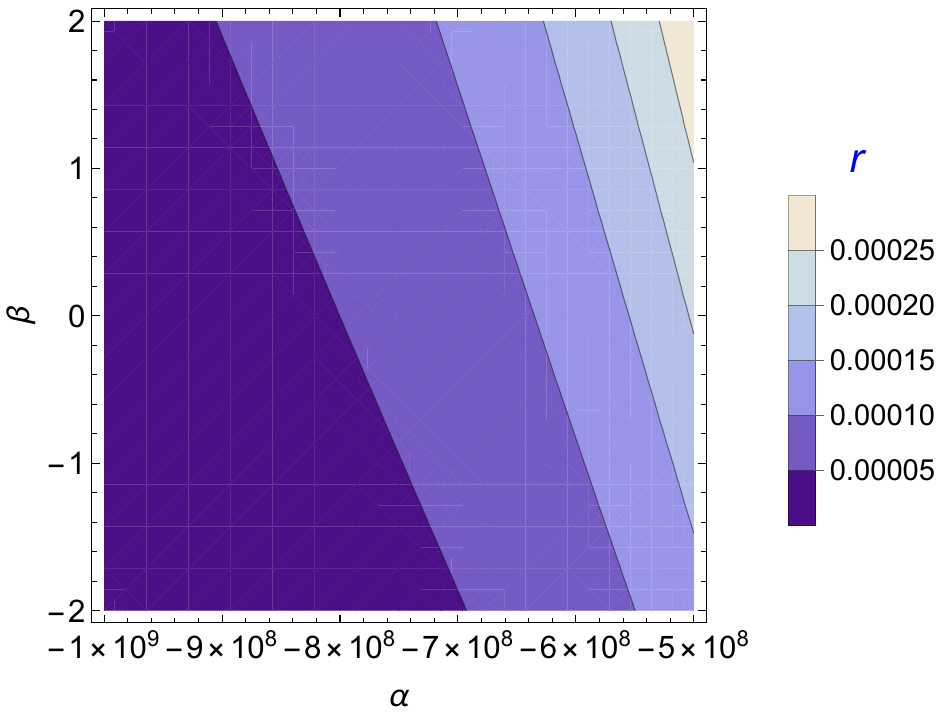}
\label{d1} }  \subfigure[] {
\includegraphics[width=0.31\textwidth]{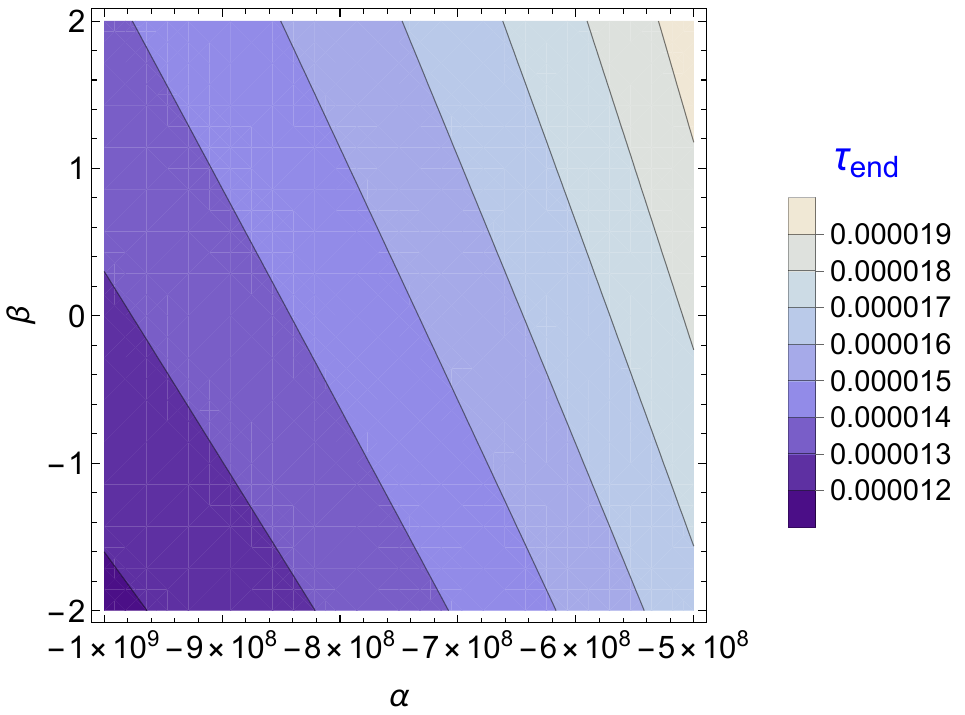}
\label{e1} }
 \caption{\label{fig1} [color online] All shapes are drawn as
functions of the model parameters $\alpha$ and $\beta$ for the
power-law potential with $n=2$, $N=70$, $\Upsilon_0= 0.5\times
10^{-8}$ and  $\nu=10^{-8}$ with $\kappa=8\pi$, wherein (a)
indicates the amplitude of the scalar power spectrum $\Delta_{\rm
R}^2$, (b) and (c) show the scalar spectral index $n_{\rm S}$, (d)
indicates the tensor-to-scalar ratio $r$, and (e) demonstrates the
temperature at the end of inflation $\tau_{\rm end}$.}
\end{figure*}

Hence, in FIG.~\ref{fig1}, we have depicted the amplitude of the
scalar power spectrum, the scalar spectral index, the
tensor-to-scalar ratio, and the temperature at the end of
inflation with respect to the model parameters $\alpha$ and
$\beta$ for the power-law potential with $n=2$, $N=70$, $\nu \sim
{\cal O}(10^{-8})$ and $\Upsilon_0\sim {\cal O}(0.5\times
10^{-8})$.\footnote{Note that in this model, since $n_{\rm S}$
does~not depend on $\nu$ and $\Upsilon_0$, by eliminating $\phi$
between the relations $r$ and $\Delta_{\rm R}^2$ and getting their
relation in terms of each other, while fixing $n=2$ and $N=70$ but
floating $\alpha$ and $\beta$, we probed to find an acceptable
value for each of $\nu$ and $\Upsilon_0$ such that the values of
$r$ and $\Delta_{\rm R}^2$ fall within the acceptable range of
observational data. Moreover, in drawing all figures, the values
chosen within the given range of $\alpha$ and $\beta$ satisfy
conditions \eqref{conseq}, \eqref{strongConditions} and
\eqref{NEWconsOnKapaBetaAlpha}.}\
 The
results indicate that, in the range of $(-10^9)<\alpha<(-0.5\times
10^9)$ and $-2< \beta<2$, the model-derived values are in good
agreement for $\Delta_{\rm R}^2$ with both the Planck $2018$ data
and the WMAP three-year data~\cite{Planck:2018jri,
Montefalcone:2022jfw} and for $r$ with both the Planck $2018$
data~(\ref{planckdata}) and the joint Planck, BK$15$ and BAO
data~\eqref{planckdata2}. Also, the model produces a variety of
spectra that span a range from blue tilt -- indicated by $n_{\rm
S}> 1$ for $(-10^9) < \alpha < (-0.5\times 10^9)$ -- to red tilt
-- indicated by $n_{\rm S} < 1$ for $-0.6 < \alpha < -0.4$ -- in
alignment with the WMAP three-year data~\cite{WMAP2006}. However,
in the range of $(-0.5\times 10^9)<\alpha<-0.6$, the spectrum has
values of $n_{\rm S}$ greater than one and less than one, although
we have~not shown the corresponding plots and have only shown the
appropriate values. Furthermore, for the positive values of
$\alpha$, the model does~not produce real results or compatible
results with the observational data. Moreover, the result of
FIG.~\ref{fig1} shows that the temperature at the end of
inflation, with the chosen values, is approximately ${\mathcal
O}(10^{-5})$. Although, by inserting the exact value of the
$\kappa $ parameter, its value becomes approximately ${\mathcal
O}(10^{28})$.

\subsection{POWER-LAW POTENTIAL WITH VARIABLE DISSIPATION COEFFICIENT}
In this subsection, we continue the calculations with the
assumption that the dissipative coefficient is
\begin{equation}\label{cubic}
\Upsilon=\Upsilon_*\left(\dfrac{\phi}{\phi_0}\right)^2,
\end{equation}
where $\Upsilon_*$ and $\phi_0$ are constant. In this case, the
slow-roll parameters can be derived in a similar way as we
obtained relations \eqref{srpc}, except for $\lambda$, which are
now
\begin{equation}\label{srpower}
\epsilon\approx \dfrac{n^2}{2\kappa\chi\phi^2}, \qquad
\eta\approx\dfrac{n(n-1)}{\kappa\chi\phi^2}, \qquad\lambda\approx
\dfrac{2n}{\kappa\chi\phi^2}.
\end{equation}
Next, from the condition $\epsilon\left(\phi_{\rm end}\right)=1$,
we can determine the value of the scalar field at the end of the
inflationary era as
\begin{equation}\label{phiend2power}
\phi_{\rm end}^2\approx \dfrac{n^2}{2\kappa\chi_{\rm end}}.
\end{equation}
Thus, we need to specify the dissipation rate at the end of the
inflationary era. For this purpose, by substituting the
dissipation coefficient \eqref{cubic} into relation
\eqref{dissipation rate}, while utilizing Eq. \eqref{h2power}, we
obtain
\begin{equation}\label{chie}
\chi \approx \sqrt{\dfrac{- \alpha \,\Upsilon_*^2
 }{3\, \nu \left(\kappa+2\beta\right)
\phi_0^4}}\,\phi^{\dfrac{\left(4-n\right)}{2}}.
\end{equation}
Meanwhile, this case with the power of $n=4$ clearly has a
constant dissipation rate. In turn, relation \eqref{phiend2power}
gives
\begin{equation}
\phi_{\rm end}\approx \left[\dfrac{3\, \nu n^4
\left(\kappa+2\beta\right)\phi_0^4}{-4 \alpha  (\kappa
\Upsilon_*)^2}\right]^{1/\left(8-n\right)},
\end{equation}
which dictates $n\neq 8$.

As in the previous subsection, the e-folding number
\eqref{Nstrong} for this case yields
\begin{align}
N\!\approx\!
\left[\phi^{\left(8-n\right)/2}\!-\!\phi^{\left(8-n\right)/2}_{\rm
end}\right]\! \left[\!\dfrac{-4 \alpha\, (\kappa \Upsilon_*)^2 }{3
\nu n^2\left(8-n\right)^2\!\left(\kappa+2\beta\right) \phi_0^4
 }\!\right]^{1/2}\! ,
\end{align}
and in turn, the scalar field at the first horizon crossing
becomes
\begin{align}\label{phicubic}
\phi \approx \left\lbrace \dfrac{-4 \alpha \(\kappa\Upsilon_*\)^2
}{3 \nu n^2 \left(\kappa+2\beta\right) \left[n^2+
\left(8-n\right)^2
N^2\right]\phi_0^4}\right\rbrace^{1/\left(n-8\right)}.
\end{align}
In addition, in this case, relation \eqref{zetastrong} is
\begin{equation}\label{zetp2}
\zeta\(\phi\)\approx
\left[\dfrac{\kappa\alpha-2\left(\kappa+\beta\right)}{\kappa+\beta+\kappa\alpha}\right]
\ln \phi.
\end{equation}

Also, by using relation \eqref{tauphidot} for this case, while
substituting relations \eqref{phidotpower} and \eqref{chie} into
it, we obtain the radiation temperature as
\begin{equation}\label{tauv}
\tau\approx\left[\dfrac{n^2}{4C_{\gamma}\kappa^2\Upsilon_*}\sqrt{\dfrac{3\,\nu^3\(\kappa
+2\beta\)^3\phi_0^4}{\(-\alpha\)^3}}\right]^{1/4}\phi^{\dfrac{3n-8}{8}}.
\end{equation}
Thereupon, by substituting relations \eqref{h2ap}, \eqref{chie},
\eqref{zetp2} and \eqref{tauv} into relation \eqref{59New} for the
strong dissipation regime, we obtain
\begin{align}
\Delta_{\rm R}^2 &\approx \sqrt{\dfrac{-n\left(\kappa+\beta+\kappa\alpha\right)}
{4 \pi^4 \kappa \phi_0}\Big[\dfrac{\Upsilon_*^2 \nu\left(\kappa+2\beta\right)^5}{48 C_{\gamma}^2 (-\alpha)^5 }\Big]^{1/4}}\nonumber\\
&~~~\times
\phi^{\dfrac{(n+32)(\kappa+\beta)+(n-16)\kappa\alpha}{8(\kappa+\beta+\kappa\alpha)}}.
\end{align}
Then, by inserting relation \eqref{phicubic} into the above
relation, the density fluctuation can be expressed in terms of the
e-folding number, $N$, and the other parameters, $n, \nu,
\Upsilon_*, \phi_0, \alpha$ and $\beta$.

Eventually, by substituting relations \eqref{srpower}, while using
relation \eqref{chie}, into relation \eqref{ssis}, we get the
scalar spectral index in terms of the inflaton field as
\begin{align}\label{nsvariable}
n_{\rm S}-1 \approx & \sqrt{\dfrac{3\,\nu
\left(\kappa+2\beta\right)\phi_0^4}{-\alpha \Upsilon_*^2}}\, n\,
\phi^{\left(n-8\right)/2}\nonumber\\
&\times\left[\dfrac{-16\left(2\kappa+2\beta-\kappa\alpha\right)-n\left(\kappa+\beta+\kappa\alpha\right)}
{8\kappa\left(\kappa+\beta+\kappa\alpha\right)}\right]\nonumber\\
\approx &
\dfrac{-16\left(2\kappa+2\beta-\kappa\alpha\right)-n\left(\kappa+\beta+\kappa\alpha\right)}
{4\left(\kappa+\beta+\kappa\alpha\right)\sqrt{n^2+
\left(8-n\right)^2N^2} },
\end{align}
where we have used relation \eqref{phicubic} for the last row.
Again, it is interesting to note that in this model for the case
of the strong dissipation regime, the scalar spectral index is
obtained only as a function of the $\alpha$, $\beta$, $n$ and $N$
parameters and independent of $\nu$ and $\Upsilon_*$. Also,
utilizing relations \eqref{dissipation rate},
\eqref{strongConditions} and \eqref{zetp2} into relation
\eqref{rstrong}, the tensor-to-scalar ratio reads
\begin{align}\label{rpv}
r\approx &\dfrac{\kappa \nu}{\tau}
\sqrt{\dfrac{H\phi_0^2\left(\kappa +2\beta\right)}{3\, \alpha
\Upsilon_* \left(\kappa
+\beta+\kappa\alpha\right)}}\coth\(\dfrac{k}{2\tau}\)\nonumber\\
&\times
\phi^{\left[\dfrac{n(\kappa+\beta+\kappa\alpha)-(5\kappa+5\beta-\kappa\alpha)
}{\kappa+\beta+\kappa\alpha}\right]}.
\end{align}
Now, inserting relations \eqref{h2power}, \eqref{phicubic} and
\eqref{tauv} into relation \eqref{rpv}, we again obtain the
inflationary observable $r$ as a function of the $\alpha$,
$\beta$, $n$, $N$, $\phi_0$, $\Upsilon_*$ and $\nu$ parameters.

\begin{figure*}[t!]
\centering \subfigure[]{
\includegraphics[width=0.31\textwidth]{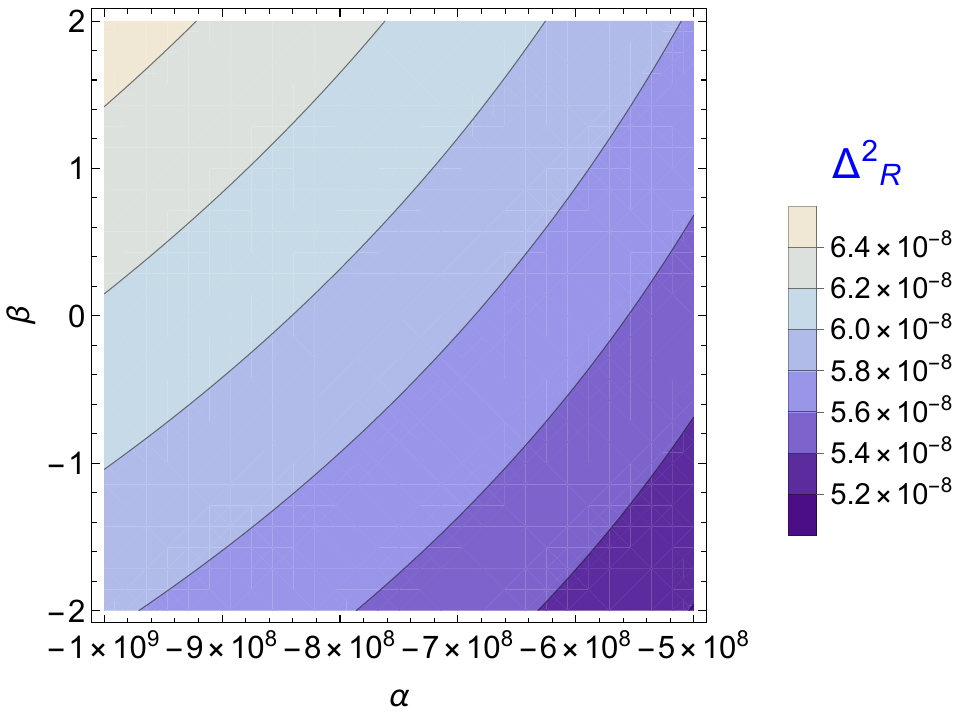}
\label{aa2} } \subfigure[] {
\includegraphics[width=0.31\textwidth]{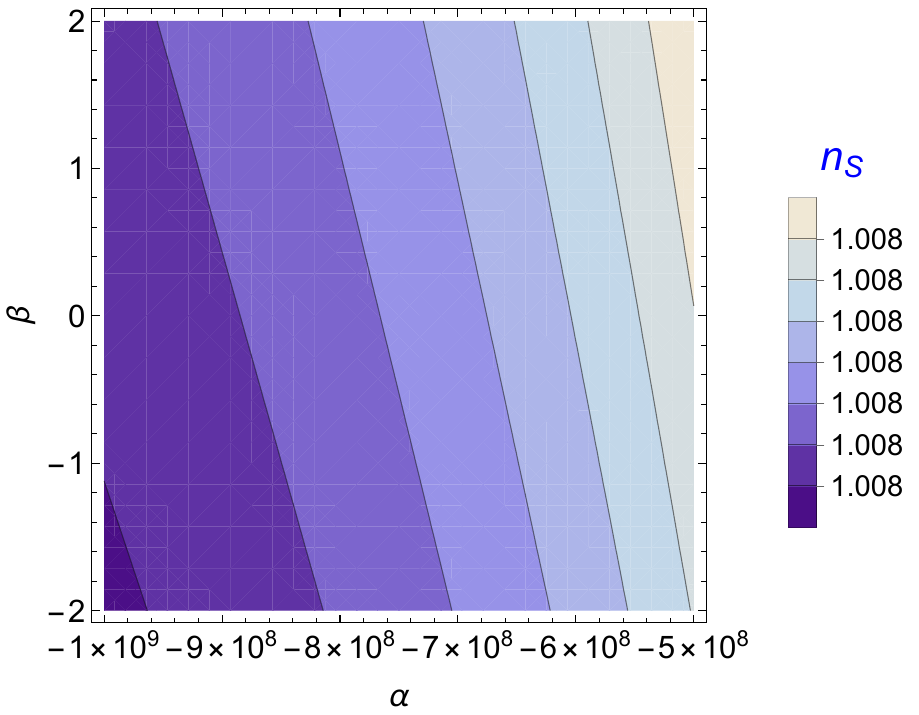}
\label{bb2} } \subfigure[]{
\includegraphics[width=0.31\textwidth]{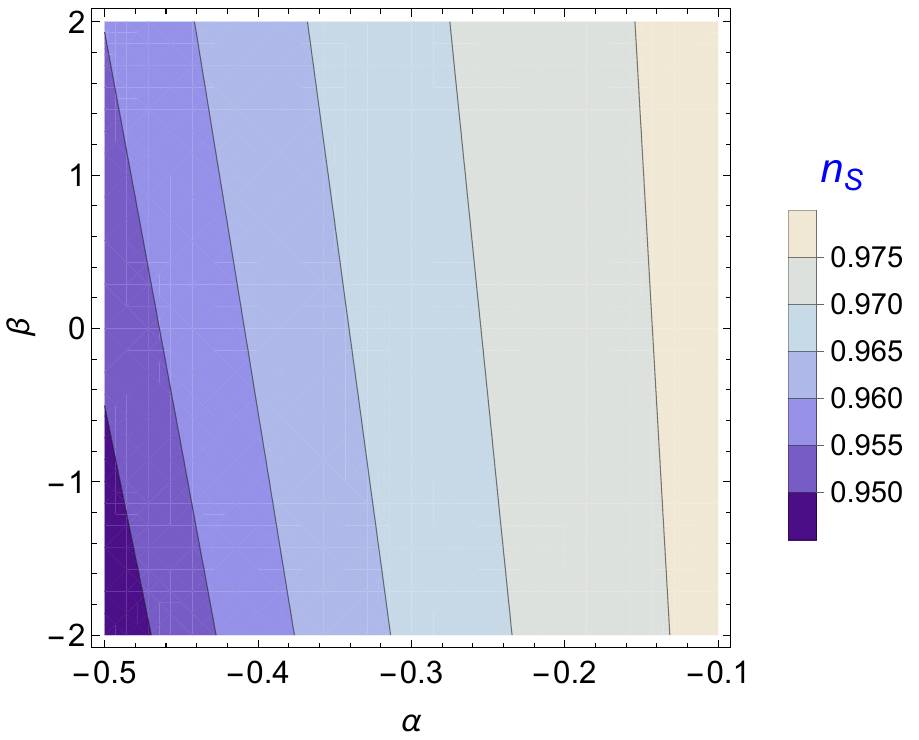}
\label{cc2} } \subfigure[] {
\includegraphics[width=0.31\textwidth]{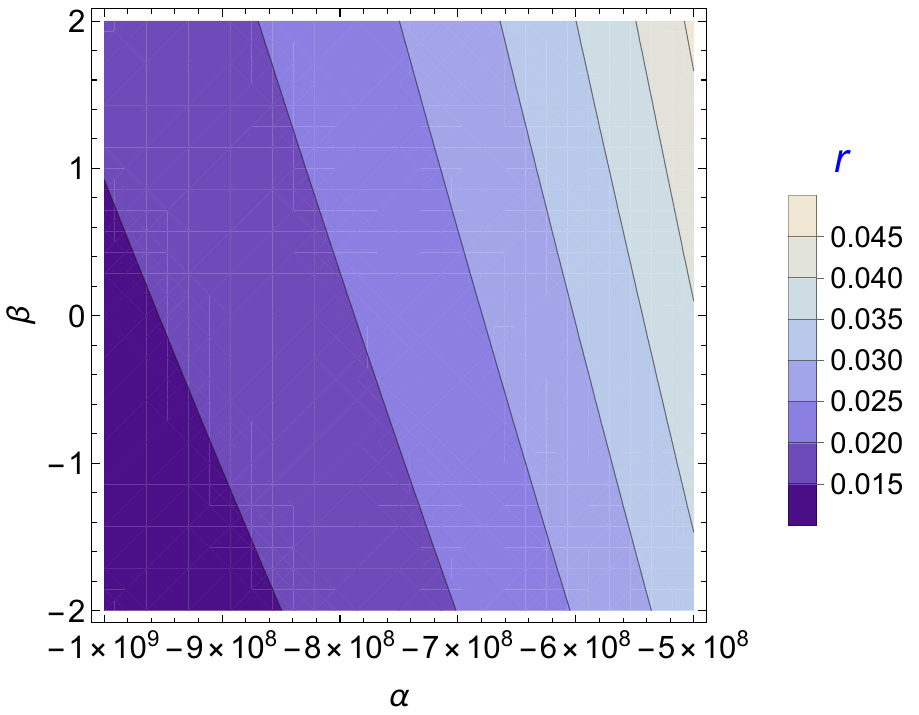}
\label{dd2} } \subfigure[] {
\includegraphics[width=0.31\textwidth]{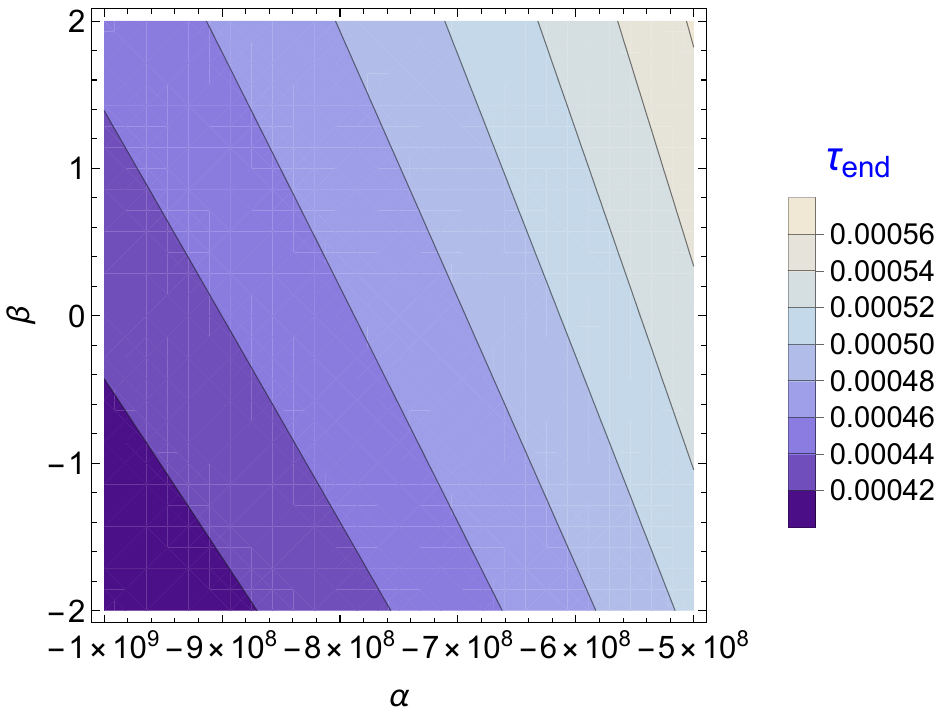}
\label{ee2} }
 \caption{\label{fig2} [color online] All shapes are drawn as
functions of the model parameters $\alpha$ and $\beta$ for the
power-law potential with $n=2$, $N=70$, $\phi_0=0.5\times 10^3$,
$\Upsilon_*= 0.5\times 10^{-3} $ and $\nu=10^{-4}$ with
$\kappa=8\pi$, wherein (a) demonstrates the amplitude of the
scalar power spectrum $\Delta_{\rm R}^2$, (b) and (c) show the
scalar spectral index $n_{\rm S}$, (d) indicates the
tensor-to-scalar ratio $r$, and (e) demonstrates the temperature
at the end of inflation $\tau_{\rm end}$. }
\end{figure*}

\begin{figure*}[t!]
\centering \subfigure[]{
\includegraphics[width=0.31\textwidth]{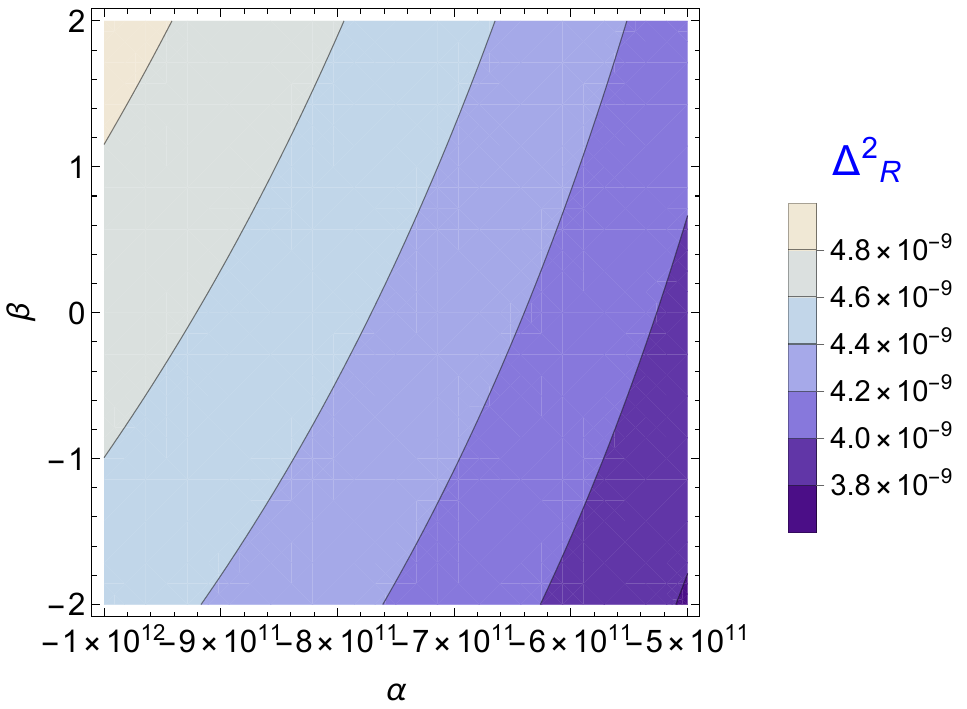}
\label{aaa3} } \subfigure[] {
\includegraphics[width=0.31\textwidth]{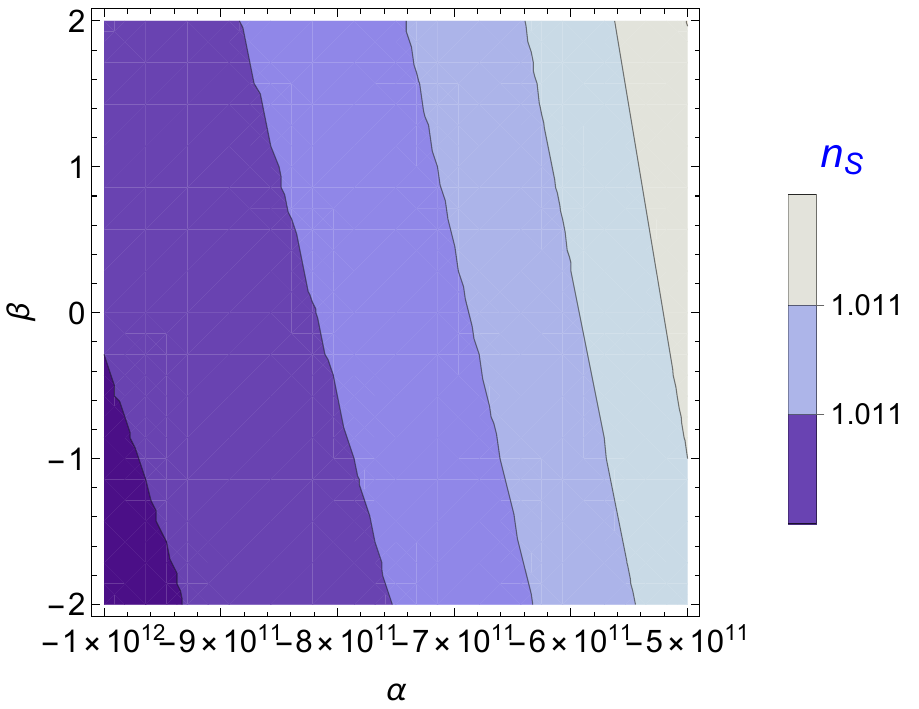}
\label{bbb3} } \subfigure[]{
\includegraphics[width=0.31\textwidth]{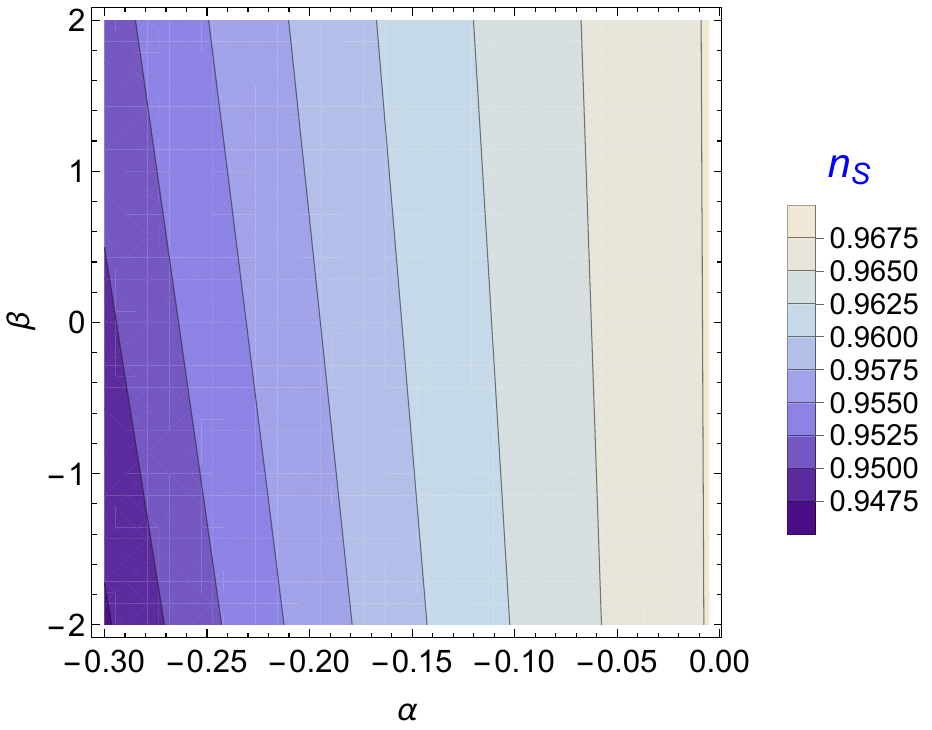}
\label{ccc3} } \subfigure[] {
\includegraphics[width=0.31\textwidth]{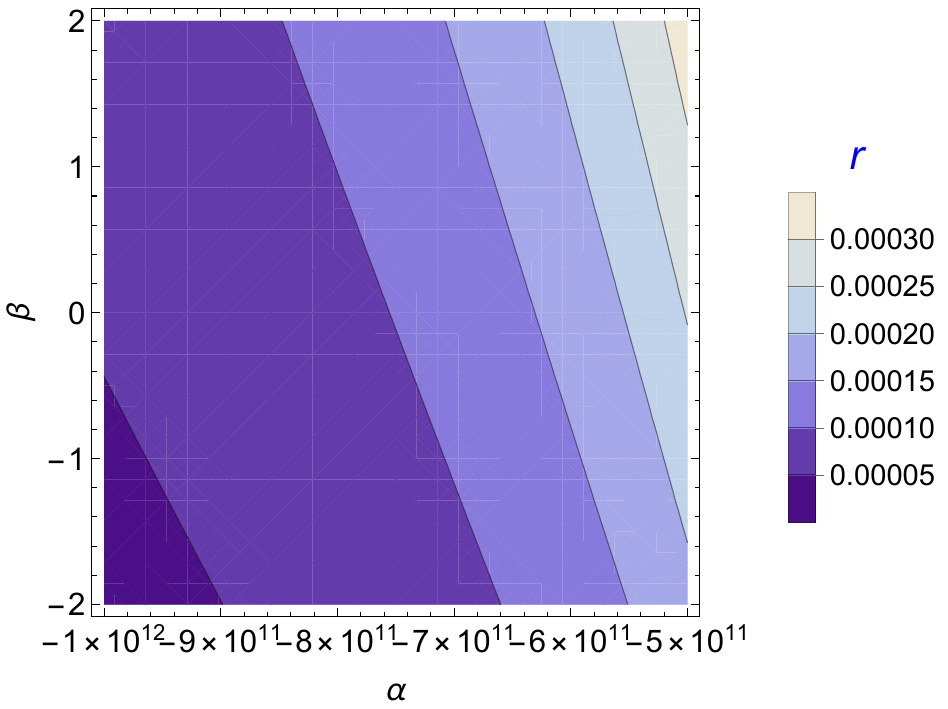}
\label{ddd3} } \subfigure[] {
\includegraphics[width=0.31\textwidth]{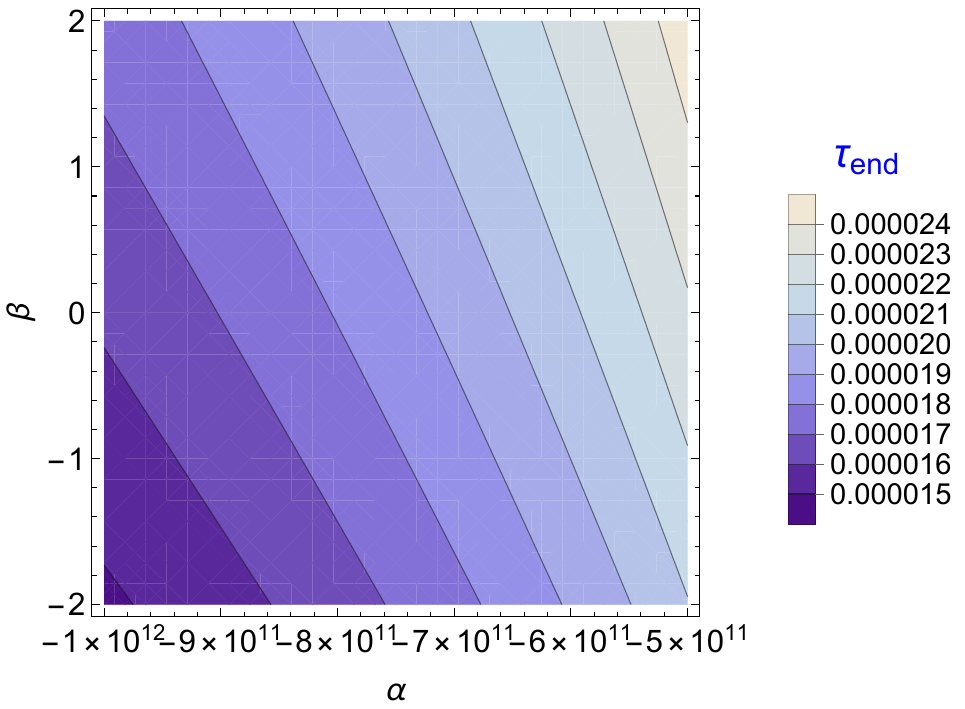}
\label{eee3} }
 \caption{\label{fig3} [color online] All shapes are drawn as
functions of the model parameters $\alpha$ and $\beta$ for the
power-law potential with $n=4$, $N=70$, $\phi_0=0.5\times 10^{4}$,
$\Upsilon_*= 10^{-7/2}$ and $\nu=10^{-8}$ with $\kappa=8\pi$,
wherein (a) demonstrates the amplitude of the scalar power
spectrum $\Delta_{\rm R}^2$, (b) and (c) show the scalar spectral
index $n_{\rm S}$, (d) indicates the tensor-to-scalar ratio $r$,
and (e) demonstrates the temperature at the end of inflation
$\tau_{\rm end}$.}
\end{figure*}

As explained in the previous subsection, we have plotted the
amplitude of the scalar power spectrum, the scalar spectral index,
the tensor-to-scalar ratio, and the temperature at the end of
inflation with respect to the model parameters $\alpha$ and
$\beta$, while fixing $N=70$, for the power-law potential with
$n=2$, $\nu \sim {\cal O}(10^{-4})$, $\Upsilon_*\sim {\cal
O}(0.5\times 10^{-3})$ and $\phi_0=0.5 \times 10^{3}$ in
FIG.~\ref{fig2}, and with $n=4$, $\nu \sim {\cal O}(10^{-8})$,
$\Upsilon_*\sim {\cal O}(10^{-7/2})$ and $\phi_0=0.5 \times
10^{4}$ in FIG.~\ref{fig3}. The results indicate that, for $n=2$
within the range of $(-10^9)<\alpha <(-0.5\times10^9)$ and
$-2<\beta <2$, the obtained values of the amplitude of the scalar
power spectrum are approximately in the range of observational
data. However, for $n=4$ within the range of $(-10^{12})<\alpha
<(-0.5\times10^{12})$ and $-2<\beta <2$, its values are more
consistent with observational data. Also, the obtained values of
$r$ for $n=2$ are approximately $\sim {\mathcal O}(10^{-2})$, and
those for $n=4$ are approximately $\sim {\mathcal
O}(10^{-5}-10^{-4})$. In both cases, the obtained values are in
good agreement with both the Planck $2018$ data~(\ref{planckdata})
and the combined Planck, BK$15$, and BAO data~(\ref{planckdata2}).
Also, the obtained results in both figures~\ref{fig2}
and~\ref{fig3} for $n_{\rm S}$ indicate that the model can
generate diverse spectra that extend across the spectrum range
from the blue tilt to the red tilt in agreement with the WMAP
three years data~\cite{WMAP2006}. However, the obtained values of
both $n_{\rm S}$ and $r$ are very sensitive to different values of
the lower bound of the $\alpha$ parameter but not~very sensitive
to different values of the parameter $\beta$. Furthermore, in
these cases, for positive values of the parameter $\alpha $,
no~real and/or compatible results are attainable. In addition, the
results show that the temperature at the end of inflation, with
the chosen values, is approximately ${\mathcal O}(10^{-4})$ in
FIG.~\ref{fig2} and ${\mathcal O}(10^{-5})$ in FIG.~\ref{fig3}.
Although, by inserting the exact value of the $\kappa $ parameter,
these values are approximately ${\mathcal O}(10^{14})$ and
${\mathcal O}(10^{13})$, respectively. Besides, for the power-law
potential with $n=4$, the dissipation rate is constant and
does~not depend on the inflaton scalar field.

In the weak dissipation regime, the scalar spectral index
\eqref{nseq} becomes
\begin{equation}
n_{\rm
S}-1=\dfrac{19}{4}\epsilon-\dfrac{5}{2}\eta-\dfrac{1}{4}\lambda.
\end{equation}
Also, relations \eqref{zetaeq} and \eqref{rstrong} are now
\begin{align}
&\zeta(\phi) \approx \int\left(\dfrac{V'}{V}-\dfrac{V''}{V'}+\chi \dfrac{\Upsilon'}{\Upsilon}\right) {\rm d}\phi,\nonumber\\
&r\approx \dfrac{\kappa
\sqrt{\left(\kappa+2\beta\right)}\,V}{3\tau \sqrt{-\alpha
\left(\kappa+\beta\right)}}\exp\left[2\zeta(\phi)\right]
\coth\left(\dfrac{k}{2\tau}\right).
\end{align}
Moreover, within this regime, the slow-roll parameters
\eqref{epsv}, \eqref{etv} and \eqref{lamv} become
\begin{align}
&\epsilon\approx\dfrac{-\alpha}{2\(\kappa+\beta\)}\(\dfrac{V^\prime}{V}\)^2,
\qquad \eta\approx\dfrac{-\alpha}
{\(\kappa+\beta\)}\(\dfrac{V^{\prime\prime}}{V}\),\nonumber\\
&\lambda\approx\dfrac{-\alpha}{\(\kappa+\beta\)}\(\dfrac{\Upsilon^
\prime V^\prime}{\Upsilon V}\).
\end{align}

\section{Conclusions}
Based on observations, it has been evident that the Universe has
experienced two periods of accelerated expansion in its history.
The first phase occurred in the early Universe, known as
inflation, and the second phase is taking place in the late
Universe, which is related to the dark energy sector.

The inflationary paradigm addresses the deficiencies of the
standard  cosmological model by introducing a period of
accelerated expansion. In general, inflationary models can be
categorized into two groups: cold inflation and warm inflation. In
the cold inflation model, the inflationary scalar field does~not
interact with other fields or radiation. It slowly rolls down
along its relatively flat potential, and the Universe experiences
a quasi-exponential expansion during inflation, which comes to an
end when the inflaton reaches the minimum of the potential. In
this case, the inflaton begins to oscillate around this minimum
and decays into the particles of the standard model, and the
Universe, which has been in a very cold phase due to very rapid
expansion, reheats and connects to the radiation-dominant era of
the standard model of cosmology. This phase is called reheating.
However, in the warm inflation model, the inflaton does interact
with other fields, leading to an energy transfer from the inflaton
to radiation field during its slow-roll. Towards the end of
inflation, the inflaton undergoes complete decay into radiation,
preventing the Universe from having a very cold phase due to
inflation. As a consequence, the Universe is seamlessly
transitioning into the radiation-dominant phase without the need
for a separate reheating phase. While clear predictions have been
successfully made by GR in describing cosmological phenomena,
these predictions are~not sufficient to fully account for the
impacts of dark sectors on the dynamics of the Universe, which are
consistent with observational data. Therefore, the exploration of
alternative gravitational theories provides a compelling
motivation to address this issue.

In this study, we have investigated warm inflation within the
framework of the linear version of $f(Q, T)$ gravity. To
accomplish this, we initially outlined the basic theoretical setup
for warm inflation, starting with GR and considering an isotropic
and homogeneous scalar field, commonly referred to as inflaton.
Next, we have provided a concise explanation of the $f(Q, T)$
gravity theory along with its mathematical aspects and
interconnections. Subsequently, we have applied this theory to a
spatially flat FLRW metric while taking into account the inflaton
scalar field and a dissipation term. As a result, we have derived
the modified Friedmann equations and the modified Klein-Gordon
equation specific to the linear version of $f(Q,T)$ gravity.
Furthermore, we have extracted the extended slow-roll parameters
and extended slow-roll conditions for this model. Following that,
we have analyzed the scalar and tensor perturbations, while we
have developed the corresponding formulations, for the linear
version of $f(Q, T)$ gravity with the warm inflation scenario. By
employing the slow-roll conditions, we have calculated both the
scalar power spectrum and the power spectrum of the tensor
perturbations. Subsequently, we have determined the scalar
spectral index and the tensor-to-scalar ratio based on these power
spectra. Then, in the strong dissipation regime, by imposing a
restriction on the parameters and assuming the power-law
potential, we have derived the amplitude of the scalar power
spectrum and the inflationary observables as well as the
temperature of the thermal bath for two cases. One case involves a
constant dissipation coefficient, while the other one considers a
more general case, a specific function of the inflaton scalar
field.

Under the assumption of the constant dissipation coefficient, we
have obtained that the scalar spectral index is a function of two
parameters of the linear model of $f(Q,T)$ gravity ($\alpha$ and
$\beta$) plus the power value $n$ of the power-law potential and
the e-folding number. Whereas, the amplitude of the scalar power
spectrum and the tensor-to-scalar ratio are functions of these
four parameters plus the constant coefficient of the potential and
the dissipation coefficient. Then, we have plotted the amplitude
of the scalar power spectrum, the scalar spectral index, the
tensor-to-scalar ratio, and the temperature at the end of
inflation with respect to the model parameters for the power-law
potential with $n=2$ while keeping the other parameters fixed. The
results obtained for the amplitude of the scalar power spectrum,
the scalar spectral index and the tensor-to-scalar ratio, in the
selected range of negative values of the $\alpha$ parameter and
positive and negative values of the $\beta$ parameter, are in good
agreement with the WMAP three years data, the Planck $2018$ data,
and the joint Planck, BK$15$ and BAO data. Also in this case, the
model generates the scalar spectral index that shifts from a blue
tilt to a red tilt as the parameter $\alpha$ increases, in
alignment with the WMAP three-year data and the Planck $2018$
data. Hence, the power-law potential can make predictions
consistent with observational data. However, these inflationary
observables rely crucially on the model parameters. For instance,
when considering positive values of the parameter $\alpha$, it
is~not possible to achieve results consistent with the
observational data. Also, the results show that the temperature of
the Universe at the end of inflation, with the chosen values, is
approximately ${\mathcal O}(10^{-5})$, although by inserting the
exact value of the $\kappa $ parameter, it is approximately
${\mathcal O}(10^{28})$.

Next, assuming that the variable dissipation coefficient is a
function of the inflaton field to the power of two, we have again
obtained that the scalar spectral index is a function of two model
parameters plus the power value of the power-law potential and the
e-folding number. Whereas, the amplitude of the scalar power
spectrum and the tensor-to-scalar ratio are functions of these
four parameters plus the constant coefficient of the potential and
the two constant coefficients of the dissipation coefficient.
Then, we have plotted the amplitude of the scalar power spectrum,
the scalar spectral index, the tensor-to-scalar ratio, and the
temperature at the end of inflation with respect to the model
parameters for the power-law potential with $n=2$ and $n=4$ while
keeping the other parameters fixed. The model produces a spectrum
for the scalar spectral index that transitions from a blue tilt to
a red tilt as the parameter $\alpha$ increases, which is in
agreement with the WMAP three years data and the Planck $2018$
data. The findings show that, for negative values of $\alpha$ and
positive and negative values of $\beta$, the results obtained for
the tensor-to-scalar ratio are consistent with both the Planck
$2018$ data and the joint Planck, BK$15$ and BAO data. For the
amplitude of the scalar power spectrum for $n=2$, they are
approximately within observational data, whereas $n=4$ matches
better with observational data. Once again, although these
inflationary observables rely crucially on the model parameters,
the power-law potential can make predictions consistent with
observational data. Besides, in this case, again, when considering
positive values of $\alpha$, it is~not possible to obtain results
in agreement with observational data. Also, the results show that
the temperature at the end of inflation, with the chosen values,
is approximately ${\mathcal O}(10^{-4})$ for $n=2$ and ${\mathcal
O}(10^{-5})$ for $n=4$, although by inserting the exact value of
the $\kappa $ parameter, these values are approximately ${\mathcal
O}(10^{14})$ and ${\mathcal O}(10^{13})$, respectively.

Finally, in contrast to the results obtained from GR in the same
scenario, since the $f(Q, T)$ gravity approach imposes tighter
constraints on the value of the tensor-to-scalar ratio, it
effectively revives the power-law potential. Such results can
provide a justification for applying $f(Q, T)$ gravity. In
addition, the utilization of the linear version of $f(Q, T)$
gravity produces distinct differences in the constraints on the
parameter of the power-law potential compared to the results
obtained under the GR assumptions. Therefore, by choosing the
appropriate parameters of the model, the linear version of
$f(Q,T)$ gravity in the warm inflationary paradigm can revive the
power-law potential.

 \vskip0.5cm

\section*{ACKNOWLEDGMENTS}
M.S. and M.F. thank the Research Council of Shahid Beheshti
University.


\end{document}